\documentclass[aps,pre,twocolumn,footinbib,longbibliography,superscriptaddress]{revtex4-1}
\usepackage{amsmath}
\usepackage{amsfonts}
\usepackage{amssymb}
\usepackage{lmodern,dsfont}
\usepackage{graphicx}
\usepackage[usenames,dvipsnames]{xcolor}
\usepackage{bm}
\usepackage{blkarray}
\usepackage{blindtext}
\usepackage[english=american]{csquotes}

\newcommand{\kb}{k_\text{B}}
\newcommand{\Av}[1]{\left\langle #1 \right\rangle}
\newcommand{\av}[1]{\langle #1 \rangle}

\newcommand{\n}{\nonumber}
\newcommand{\nn}{\nonumber \\}

\newcommand{\lrbra}[1]{\left[ #1 \right]}

\begin{document}

\author{Sosuke Ito}
\affiliation{Universal Biology Institute, the University of Tokyo,  7-3-1 Hongo Bunkyo-ku Hongo, Tokyo 113-0033, Japan}
\affiliation{JST, PRESTO, 4-1-8 Honcho, Kawaguchi, Saitama, 332-0012, Japan}
\author{Andreas Dechant}
\affiliation{WPI-Advanced Institute of Materials Research (WPI-AIMR), Tohoku University, Sendai 980-8577, Japan}
\title{Stochastic time-evolution, information geometry and the Cram{\'e}r-Rao bound}
\date{\today}

\begin{abstract}
We investigate the connection between the time-evolution of averages of stochastic quantities and the Fisher information and its induced statistical length.
As a consequence of the Cram{\'e}r-Rao bound, we find that the rate of change of the average of any observable is bounded from above by its variance times the temporal Fisher information.
As a consequence of this bound, we obtain a speed limit on the evolution of stochastic observables: Changing the average of an observable requires a minimum amount of time given by the change in the average squared, divided by the fluctuations of the observable times the thermodynamic cost of the transformation.
In particular for relaxation dynamics, which do not depend on time explicitly, we show that the Fisher information is a monotonically decreasing function of time and that this minimal required time is determined by the initial preparation of the system.
We further show that the monotonicity of the Fisher information can be used to detect hidden variables in the system and demonstrate our findings for simple examples of continuous and discrete random processes.
\end{abstract}

\maketitle

\section{Introduction} \label{sec-intro}

Information geometry \cite{Ama07} is a branch of information theory which describes information in terms of differential geometry.
This can be motivated by a question central to any physical experiment: Given a system described by a set of parameters, how much information about the system can we gain by observing its change under a variation of the parameters?
As it turns out, the effect of smooth parameter variations defines a metric called Fisher information metric \cite{Uff99,Fri04,Cro07,Cov12}.
This metric encodes the maximum amount of information that can be gained by measuring the change of any observable due to the parameter change.

The relation between the measurement of observables and information gained about the physical system is also central to thermodynamics.
Deciding which observables to measure and which parameters to vary in doing so is essential for reconstructing the thermodynamic potentials and thus obtaining complete information about the macroscopic state of the system.
Thus, it is not surprising that there exists a strong connection between thermodynamics and information theory, which, despite dating back all the way to Gibbs and Boltzmann \cite{Jay65}, has recently received much attention \cite{Kaw07,All09,Sti12,Sag12,Ito13,Har14,Hor14,Mac15,Lah16,Dec18B}.
This is in part motivated by improved experimental techniques, allowing to probe the relation between information and thermodynamic quantities on a more detailed and microscopic level \cite{Ber12,Gav17}, but also by new theoretical proposals based on understanding information as a quantity that is just as physical as matter or energy.

Interpreting thermodynamics in terms of information relies on relating mathematical measures of information to measurable physical quantities.
A fruitful approach is that of stochastic thermodynamics, which describes the behavior of thermodynamic quantities like heat, work and entropy in small systems, where these quantities fluctuate due to the presence of noise \cite{Sek10,Sei12}.
The probabilistic nature of the description allows to explicitly compute different measures of information, whose mathematical properties can then be used to make predictions about physical quantities.
For example, a central result of stochastic thermodynamics is the identification of the Kullback-Leibler divergence between the path measures of a stochastic evolution and its time-reverse with thermodynamic entropy production \cite{Sei12}.
This identification allows for an immediate proof of the fluctuation theorem and also guarantees that the entropy production is always positive.
More recently, Ref.~\cite{Ito17} found an intimate connection between stochastic entropy and statistical length, relating stochastic thermodynamics and information geometry.
As a consequence of this relation, a universal speed limit for stochastic dynamics was obtained.

In this work, we focus on the physical interpretation of Fisher information and its consequences on the time-evolution of stochastic systems and observables.
The Fisher information measures how much information the state of the system contains about a set of parameters.
If the state of the system does not depend on one of the parameters, then the associated Fisher information is zero and no information about this parameter can be gained from a measurement.
On the other hand, if a small variation of a parameter causes a large change in the system's state, then the Fisher information is large and we can obtain a precise estimate of the parameter using a suitable measurement.
In the typical setting, the state of the system is described by a set of probabilities or a probability density, which depend on external parameters like temperature, pressure or external fields.

Here, we take an entirely different viewpoint: instead of external parameters, we consider time itself as a parameter.
As our first main result, we argue that the Fisher information can be interpreted as an intrinsic evolution speed of the system.
This intrinsic speed has two important properties:
First, it can be obtained by only measuring the evolution of the state of the system and does not require any information about the microscopic details of the dynamics.
Second, and more importantly, the intrinsic speed defined by the Fisher information limits the evolution speed of arbitrary observables.
More specifically, the rate of change of any observable is smaller than the product of its fluctuations and the Fisher information.
This speed limit is a consequence of the Cram{\'e}r-Rao bound \cite{Rao45,Cra16}, a central result of estimation theory.
Further interpreting the Fisher information as a thermodynamic cost \cite{Ito17}, this result complements a class of recently derived steady-state thermodynamic uncertainty relations \cite{Bar15,Gin16,Pie17,Hor17,Dec17,Mae17}.
In the particular case of an equilibrium system with quasistatic driving \cite{Cro07}, we show that the Fisher information and thus the intrinsic speed is determined by the fluctuations of the driving power.

As our second main result, we show that for Markovian relaxation processes, the Fisher information is a monotonically decreasing function of time.
This finding quantifies the intuitive expectation that the dynamics during a relaxation process should gradually slow down.
Together with the speed limit, this implies that any observable either relaxes monotonically towards its steady state value, or, if it exhibits oscillations during the relaxation, the amplitude of the oscillations has to decay.
The monotonicity of the Fisher information for relaxation processes has two further profound consequences: First, it results in a lower bound on the time required to relax a stochastic system from an initial to a final configuration, extending previously obtained speed limits for stochastic dynamics \cite{Ito17,Oku18}. 
Second, it can serve as an indicator for the presence of hidden variables in the system:
If we observe an increase of the Fisher information during a relaxation process, this necessarily implies that we are missing some information about the system.
We show that this discrepancy between observed and total information can be used to detect hidden degrees of freedom.

\section{Intrinsic speed of stochastic dynamics} \label{sec-intrinsic-speed}

Throughout the paper, we consider a system that can be described by a set of stochastic, time-dependent quantities $\bm{x}(t) = (x_1(t), \ldots x_M(t))$.
We assume that these take continuous values in $\mathbb{R}$; the case of discrete-valued processes is explicitly discussed in the Appendix.
Then, the dynamics of the system can be described in terms of a time-dependent probability density $P(\bm{x},t) \equiv P(\bm{x}(t) = \bm{x})$.
Examples for $\bm{x}(t)$ are the positions and/or velocities of diffusing particles or the angles of a set of coupled rotors in the presence of noise.
Generally, the probability density $P(\bm{x},t)$ will change as a function of time, either due to the presence of time-dependent forces in the system or due to relaxation from a given initial state $P(\bm{x},0)$.
We are further interested in an observable $R(\bm{x}(t))$, which is a function of $\bm{x}(t)$ and could be the center of mass of a particle system or its total energy.
By definition, the average of this observable is expressed in terms of the probability density as $\av{r}_t = \int d\bm{x} \ r(\bm{x}) P(\bm{x},t)$, which is the expected value of $r$ for all possible realizations of $\bm{x}(t)$.
However, since the $\bm{x}(t)$ are stochastic quantities, the observable $r(\bm{x}(t))$ is generally a fluctuating quantity.
The fluctuations of $r$ can be quantified in terms of its variance $\av{\Delta r^2}_t = \av{r^2}_t - \av{r}_t^2$, which tells us how much we expect $r$ to deviate from its average for any single realization of $\bm{x}(t)$.
As the system evolves in time, so will the average and variance of the observable $r$.
Our first main result is based on the Cram{\'e}r-Rao bound \cite{Rao45,Cra16}: the variance of an estimator of a parameter $\theta$ is always larger than one over the Fisher information. 
Choosing the parameter $\theta$ as time, the Cram{\'e}r-Rao bound can be interpreted as a speed limit on the evolution of an arbitrary observable,
\begin{align}
\big\vert d_t \av{r}_t \big\vert \leq \sqrt{\av{\Delta r^2}_t} v_I(t) \label{speed-limit-0} ,
\end{align}
where $v_I(t) = \sqrt{I(t)}$ is the square root of the Fisher information
\begin{align}
I(t) = \int d\bm{x} \ \frac{\big(\partial_t P(\bm{x},t) \big)^2}{P(\bm{x},t)} \label{fisher} .
\end{align}
In the sense of the Cram{\'e}r-Rao bound, we use a time-dependent observable $\av{r}_t$ as an estimator of $t$ and thus its fluctuations are bounded from below by $1/I(t)$.
Crucially, the speed $v_I(t)$ is independent of the choice of $r$ and thus can be interpreted as the intrinsic speed of the evolution.
Thus the speed limit Eq.~\eqref{speed-limit-0} can be understood in the following way:
The Fisher information determines the intrinsic speed $v_I(t)$ with which the system evolves.
This intrinsic speed also limits the rate at which any observable, measured in units of its typical fluctuations, can change.
We stress that that $v_I(t)$ depends only on the measurable time evolution of the probability density; it does not require any knowledge about the microscopic dynamics.

Another way to understand Eq.~\eqref{speed-limit-0} is in terms of the typical scales involved in the time evolution of an observable:
We require at least two scales to characterize this evolution.
One is the scale on which we measure the observable itself, e.~g.~length, energy or momentum, depending on the observable.
The other is the time scale on which the expectation of the observable changes.
Eq.~\eqref{speed-limit-0} suggests that a natural scale for the observable is set by its root-mean-square fluctuations $\bar{r} = \sqrt{\av{\Delta r^2}}$.
When measuring the rate of change of the observable relative to this scale, this naturally yields the time scale $\tau_r = \bar{r}/\vert d_t \av{r}_t \vert$.
Importantly, these scales can be obtained solely from a measurement of $r(t)$ and do not require any modeling of the dynamics.
Equation \eqref{speed-limit-0} then ensures that the time scale governing the evolution of any observable can only be larger than the intrinsic time scale set by the Fisher information.

%The speed $v_I(t)$ inherits a useful property from the Fisher information (see also Section \ref{sec-fisher-def}):
%Suppose that we divide the stochastic observables $\bm{x}(t)$ into two sets $\bm{x}(t) = (\bm{y}(t),\bm{\psi}(t))$.
%For example, $\bm{\psi}(t)$ may be fast degrees of freedom, while $\bm{y}(t)$ are the slow, observable degrees of freedom.
%Then the speed $v_{I,y}(t)$ defined by the distribution $P_y(\bm{y},t) = \int d\bm{\psi} \ P(\bm{y},\bm{\psi},t)$ is always less than the original speed, $v_{I,y}(t) \leq v_I(t)$.
%As a consequence, any observable which only depends on $\bm{y}(t)$ can at most evolve at the slower rate set by $v_{I,y}(t)$.
%Conversely, any observable that changes faster than that has to depend some of the fast degrees of freedom $\bm{\psi}(t)$.
%We elaborate on this observation in Section \ref{sec-monotonicity}.

An obvious question is if, and under what conditions, we can have equality in Eq.~\eqref{speed-limit-0}, i.~e.~whether some observable can change as fast as is allowed by the change of the probability density.
As we show in Appendix \ref{app-sufficient}, this can only be realized if the probability distribution can be written as
\begin{align}
P(\bm{x},t) = \frac{e^{A(t) r(\bm{x})}}{\av{e^{A(t) r}}_0} P_0(\bm{x}) 
\end{align}
where the function $A(t)$ can be expressed in terms of the average and variance of $r$ as
\begin{align}
A(t) = \int_0^t ds \ \frac{d_s \av{r}_s}{\av{\Delta r^2}_s} .
\end{align}
This form of the probability distribution means that, starting from an arbitrary initial distribution, the time evolution can be described as exponentially tilting the distribution by the observable $r(\bm{x})$.
For any probability distribution that cannot be written in the above form, the speed limit Eq.~\eqref{speed-limit-0} is an inequality.

As an example of a stochastic dynamics, we consider a particle system in contact with a heat bath described by the Langevin equation
\begin{align}
\dot{\bm{x}}(t) = \bm{\mu} \bm{f}(\bm{x}(t),t) + \sqrt{2 \bm{\mu} \kb T} \bm{\xi}(t) \label{langevin-0} .
\end{align}
Here $\bm{\mu}$ is the positive definite and symmetric mobility matrix, $\bm{f}$ contains the systematic forces (interactions and external forces) acting on the particles, $\kb$ is the Boltzmann constant and $T$ the temperature.
The vector $\bm{\xi}$ is composed of independent Gaussian white noises $\av{\xi_i(t) \xi_j(s)} = \delta_{i j} \delta(t-s)$.
Equivalently, we can describe the system in terms of its probability distribution $P(\bm{x},t)$, which evolves according to the Fokker-Planck equation
\begin{subequations}
\begin{align}
\partial_t P(\bm{x},t) &= - \bm{\nabla} \cdot \bm{j}(\bm{x},t)  \\
\bm{j}(\bm{x},t) &= \bm{\mu} \big( \bm{f}(\bm{x},t) - \kb T \bm{\nabla} \big) P(\bm{x},t) ,
\end{align} \label{fokkerplanck}%
\end{subequations}
where $\bm{j}$ is the probability current.
Depending on $\bm{f}$, the dynamics can have several time scales: the relaxation of the particles in an external potential, interactions between the particles and explicitly time-dependent forces.
Further, different observables may depend on these time scales in different ways.
For example, the center of mass position may be insensitive to interactions between the particles and thus not vary on the corresponding time scale.
On the other hand, the relative distance between particles is generally sensitive to interactions and may vary considerably.
Nevertheless, both observables obey the speed limit \eqref{speed-limit-0} and thus the the time scale $1/v_I(t)$ set by the Fisher information dominates all other time scales in the system.

A bound resembling Eq.~\eqref{speed-limit-0} may also be obtained in terms of the entropy production rate \cite{Che06,Spi12}
\begin{align}
\sigma^\text{tot}(t) = \frac{1}{\kb T} \int d\bm{x} \ \frac{\bm{j}(\bm{x},t)^T \bm{\mu}^{-1} \bm{j}(\bm{x},t)}{P(\bm{x},t)} \label{entropy} ,
\end{align}
where the superscript $T$ denotes transposition.
As shown in Ref.~\cite{Dec18}, we then have the inequality
\begin{align}
\big\vert d_t \av{r}_t \big\vert &\leq \sqrt{\chi_{r}(t)} \ \sqrt{\sigma^\text{tot}(t)} \label{entropy-speed-limit} \\
\text{with} \quad \chi_{r}(t) &= \kb T \int d\bm{x} \ \bm{\nabla} r(\bm{x})^T \bm{\mu} \bm{\nabla} r(\bm{x}) P(\bm{x},t) \n .
\end{align}
While this inequality shows that any time evolution necessarily involves a finite entropy production rate and can be considered as a kind of entropic speed limit, it does not lend itself to a straightforward interpretation in terms of an intrinsic speed of the dynamics.
First of all, the quantity $\chi_r$ has dimensions of $r^2/\text{time}$, resembling a diffusion coefficient for $r$ and thus contains a timescale by itself.
As a consequence, the corresponding speed limit is a combination of the stochastic change of $r$ over small length scales (as signified by the appearance of the gradient in $\chi_r$) and the rate of entropy production.
Second, the entropy production rate itself depends on the microscopic dynamics, i.~e.~we generally need to know the forces and fluxes in the system to calculate it.
By contrast, the speed $v_I(t)$ in Eq.~\eqref{speed-limit-0}, which is defined in terms of the Fisher information, only depends on the time-evolution of the probability density.
The quadratic form of the Fisher information Eq.~\eqref{fisher} and entropy production rate Eq.~\eqref{entropy} allows both quantities to be interpreted as a kinetic energy of the stochastic dynamics.
Specifically, defining $\dot{\Phi}(\bm{x},t) = - \partial_t \ln (P(\bm{x},t))$ and $\bm{\nu}(\bm{x},t) = \bm{j}(\bm{x},t)/P(\bm{x},t)$ we can write
\begin{subequations}
\begin{align}
I(t) &= \av{\dot{\Phi}^2}_t \\
\sigma^\text{tot}(t) &= \frac{1}{\kb T} \av{\bm{\nu}^T \bm{\mu}^{-1} \bm{\nu}}_t.
\end{align}
\end{subequations}
The local mean velocity $\bm{\nu}(\bm{x},t)$ describes the average flow at coordinate $\bm{x}$ and time $t$.
It vanishes in equilibrium and takes a finite, time-independent value in a non-equilibrium steady state.
Thus, while $\dot{\Phi}(\bm{x},t)$, the stochastic rate of Shannon entropy production, describes the rate of change of the probability distribution, the local mean velocity $\bm{\nu}(\bm{x},t)$ describes how the stochastic degrees of freedom $\bm{x}(t)$ themselves change with time.
The Fisher information represents the kinetic energy associated with the evolution of the probability distribution, while the entropy production is the kinetic energy of the local flows.
As remarked above, the entropy production contains more information about the microscopic dynamics.
Similarly to Eq.~\eqref{entropy-speed-limit} we have
\begin{align}
I(t)^2 \leq \kb T \av{\bm{\nabla} \dot{\Phi}^T \bm{\mu} \bm{\nabla} \dot{\Phi}} \ \sigma^\text{tot}(t) \label{entropy-fisher}
\end{align}
and thus a vanishing entropy production also implies a vanishing Fisher information, while the converse is true only for systems that relax to an equilibrium state.
We remark that generally, either one of the bounds Eqs.~\eqref{speed-limit-0} or \eqref{entropy-speed-limit} may be tighter.

Finally let us remark on the connection between Eq.~\eqref{speed-limit-0} and the thermodynamic uncertainty relation \cite{Bar15,Gin16}, another inequality that has recently received much attention \cite{Pie17,Hor17,Dec17}.
While the above discussion focuses on observables in the usual sense, i.~e.~whose average can be expressed in terms of the probability density, another class of observables, so-called time-integrated currents, can be defined in terms of the local mean velocity as
\begin{align}
\av{q}_t \equiv \int_0^t dt' \int d\bm{x} \ \bm{\rho}(\bm{x}) \bm{\nu}(\bm{x},t') P(\bm{x},t') .
\end{align}
In contrast to usual observables, such currents can change as a function of time even the probability density does not change in time, provided that the local mean velocity is non-zero.
In terms of the Fokker-Planck equation \eqref{fokkerplanck}, this occurs if the probability current is non-zero but its divergence vanishes and is referred to as a non-equilibrium steady state (in contrast to a true equilibrium state, where the current is zero).
For such a non-equilibrium steady state, the thermodynamic uncertainty relation \cite{Bar15,Gin16} states that the rate of change of the currents is bounded from above by the rate of entropy production,
\begin{align}
\big\vert d_t \av{q}^\text{st}_t \big\vert \leq \sqrt{D_q} \sqrt{\sigma^\text{tot,st}} \label{uncertainty} ,
\end{align}
where $D_q = \lim_{t \rightarrow \infty} \av{\Delta q^2}_t/(2 t)$ is the diffusion coefficient associated with the current and the superscript \enquote{st} denotes the steady state.
Equations~\eqref{speed-limit-0} and \eqref{uncertainty} are similar in that they provide a bound on the rate of change of an observable in terms of its fluctuations and a positive, information theoretic quantity.
However, they apply to complementary physical situations.
In the thermodynamic uncertainty relation \eqref{uncertainty}, the system is in a steady state and the observable is a current, i.~e.~depends on transitions in the system.
This non-equilibrium steady state is characterized by an increase in entropy, which by Eq.~\eqref{entropy} can be interpreted as the magnitude of the intrinsic currents.
On the other hand, the speed limit Eq.~\eqref{speed-limit-0} describes a system with an explicit time evolution and the observable depends only on the current state of the system rather than transitions.
The rate of change of such observables vanishes in the steady state and is governed by the Fisher information rather than the entropy production.

\section{Monotonicity of relaxation processes} \label{sec-monotonicity}
From Eq.~\eqref{speed-limit-0}, it is clear that the intrinsic velocity $v_I(t)$ is generally a dynamical quantity rather than just a parameter; in particular, it depends explicitly on time.
It is thus a natural question how this velocity and thus the speed of the time evolution changes over its course.
To answer this question, we note that since $d_t v_I(t) = d_t I(t)/(2 \sqrt{I(t)})$, and $I(t)$ is positive, the velocity inherits the qualitative dynamics of the Fisher information.
For a Fokker-Planck dynamics like Eq.~\eqref{fokkerplanck}, we show in Appendix \ref{app-fisher-markov} that the time derivative of the Fisher information can be decomposed into two parts,
\begin{align}
d_t I(t) = d_t I_\text{drv}(t) + d_t I_\text{rel}(t) \label{fisher-split},
\end{align}
which are explicitly given by
\begin{subequations}
\begin{align}
d_t I_\text{drv}(t) &= -2 \int d\bm{x} \ \dot{\bm{f}}(\bm{x},t)^T \bm{\mu} \bm{\nabla}\dot{\Phi}(\bm{x},t) P(\bm{x},t) \\
d_t I_\text{rel}(t) &= -2 \kb T \int d\bm{x} \ \bm{\nabla}\dot{\Phi}(\bm{x},t)^T \bm{\mu} \bm{\nabla}\dot{\Phi}(\bm{x},t) P(\bm{x},t) .
\end{align}\label{fisher-split-2}%
\end{subequations}
Here we used the notation $\dot{\bm{f}}(\bm{x},t) = \partial_t \bm{f}(\bm{x},t)$ and $\Phi(\bm{x},t) = - \ln(P(\bm{x},t))$ as above.
The first term $d_t I_\text{drv}(t)$ explicitly contains the time derivative of the force.
We refer to this term as the driving term, which only appears when we apply a time-dependent driving to the system.
We note that this term can be positive or negative.
By contrast, the second term is always negative since the mobility matrix is positive definite.
Thus this term always decreases the Fisher information and we interpret is as the relaxation of the system towards the instantaneous steady state.
In particular, if there is no time-dependent driving, then the Fisher information is a monotonically decreasing function of time,
\begin{align}
d_t I(t) = d_t I_\text{rel}(t) \leq 0 \label{fisher-monotonic} .
\end{align}
As we show in Appendix \ref{app-fisher-markov} this property is not specific to dynamics of the form Eq.~\eqref{fokkerplanck} but holds for general Markovian dynamics without explicit time dependence.
We thus have the general statement:
For a Markovian relaxation process, the Fisher information is a monotonically decreasing function of time.

Since the same statement holds for the velocity $v_I(t)$, the intrinsic speed of a relaxation process monotonically decreases with time.
Further, as a consequence of the speed limit Eq.~\eqref{speed-limit-0}, the rate of change of any observable is bounded by a monotonically decreasing function of time, which implies that also the evolution of observables gradually slows down during a relaxation process.
Note that this does not imply that the relaxation of arbitrary observables is monotonic; in the general case there may be oscillations even during a purely relaxational dynamics, however, as a consequence of the speed limit, the amplitude of these oscillations has to decrease with time.

A general property of the Fisher information (see Section~\ref{sec-fisher-def} below) is its additivity under a separation of variables.
Suppose that, as in Eq.~\eqref{fisher-add}, the system of interest is composed of two sets of degrees of freedom $\bm{y}$ and $\bm{\psi}$.
Physically, we assume that $\bm{y}$ contains the observable degrees of freedom, that are accessible to direct observation, and $\bm{\psi}$ is composed of hidden degrees of freedom, which are not directly observable.
The Fisher information is additive under this separation of variables (see Eq.~\eqref{fisher-add}),
\begin{align}
I(t) = I_{\psi \vert y}(t) + I_y(t),
\end{align}
where $I_y(t)$ is the Fisher information corresponding to the probability density of the observable variables $P_y(\bm{y},t)$.
If the system is time-independent, we then have for the Fisher information of the total system from Eq.~\eqref{fisher-monotonic}
\begin{align}
d_t I(t) = d_t I_{\psi \vert y}(t) + d_t I_y(t) \leq 0 .
\end{align}
While each individual term may be positive or negative, the sum of the terms has to be negative. 
This means that if we measure $I_y(t)$ from the probability distribution of the observable degrees of freedom and find $d_t I_y(t) > 0$ at any time, than this is a clear indicator that hidden degrees of freedom are present in the system.
We can make this precise in the form of the following statement:
If, for some stochastic process $\bm{y}(t)$, we observe $d_t I_y(t) > 0$ at any time $t$, the process cannot be described in terms of a diffusion process with time-independent drift and diffusion coefficient.
Thus, either the drift and/or diffusion coefficient depend explicitly on time, or there are hidden degrees of freedom in the system which effectively render the process $\bm{y}(t)$ non-Markovian.
We stress that this criterion relies only on a measurement of the probability density $P_y(\bm{y},t)$ of the observable degrees of freedom.
It thus provides a way to detect hidden degrees of freedom directly from a measurement, without assuming any kind of model for the dynamics.
Further, while a mismatch between the measured probability density and a given model generally only implies that the specific model cannot explain the experiment, a violation of Eq.~\eqref{fisher-monotonic} rules out \emph{any} model with the same number (or less) of degrees of freedom.

For dynamics which are driven by time-dependent forces, the Fisher information can decrease or increase as a function of time.
Still, the decomposition of its derivative Eq.~\eqref{fisher-split-2} into the negative relaxation and the (positive or negative) driving part remains valid.
Further, the two contributions are not independent.
Rather, applying the Cauchy-Schwarz inequality to the driving part, we obtain
\begin{align}
\big(d_t I_\text{drv}(t) \big)^2 \leq 4 \Av{\dot{\bm{f}}^T \bm{\mu} \dot{\bm{f}}}_t \Av{\bm{\nabla} \dot{\Phi}^T \bm{\mu} \bm{\nabla} \dot{\Phi}}_t.
\end{align}
Identifying the second term on the right-hand side as the relaxation part, this can be written as
\begin{align}
\big(d_t I_\text{drv}(t) \big)^2 \leq - \frac{2}{\kb T} \Av{\dot{\bm{f}}^T \bm{\mu} \dot{\bm{f}}}_t d_t I_\text{rel}(t).
\end{align}
The first term on the right-hand side only depends on the time-derivative of the applied forces.
Thus, the maximal magnitude of the driving part is bounded by the relaxation part.
Plugging in Eq.~\eqref{fisher-split-2}, we obtain
\begin{align}
\big(d_t I_\text{drv}(t) \big)^2 &- \frac{2}{\kb T} \Av{\dot{\bm{f}}^T \bm{\mu} \dot{\bm{f}}}_t d_t I_\text{drv}(t) \\ 
& \quad + \frac{2}{\kb T} \Av{\dot{\bm{f}}^T \bm{\mu} \dot{\bm{f}}}_t d_t I(t) \leq 0 \n .
\end{align}
The left-hand side is a parabola in $d_t I_\text{drv}(t)$ with a positive coefficient in front of the quadratic term.
Thus, it can only be negative if the discriminant is positive, which leads to the condition
\begin{align}
d_t I(t) \leq \frac{1}{2 \kb T} \Av{\dot{\bm{f}}^T \bm{\mu} \dot{\bm{f}}}_t .
\end{align}
This provides an upper bound on the change in Fisher information in terms of the time derivative of the applied forces.
While this inequality is typically not very sharp, it allows us to estimate how much the speed of the dynamics can increase by applying a time-dependent driving to the system. 
In particular, if the forces have no explicit time dependence, the right-hand side is zero and we recover Eq.~\eqref{fisher-monotonic}.

\section{Thermodynamic interpretation of Fisher information}

The Fisher information can be explicitly expressed in terms of the energetics of the system if the probability density belongs to the exponential family \cite{Cro07},
\begin{align}
P(\bm{x},t) = \frac{e^{- \beta H(\bm{x},t)}}{\int d\bm{x} \ e^{-\beta H(\bm{x},t)}} \label{exponential-familiy},
\end{align}
where $H(\bm{x},t)$ is the Hamiltonian of the generating dynamics and $\beta = 1/(\kb T)$.
This form is realized, for example, for $\bm{f}(\bm{x},t) = -\bm{\nabla}U(\bm{x},t)$ in Eq.~\eqref{fokkerplanck}, where $U(\bm{x},t)$ is a potential that varies slowly in time.
In the quasistatic limit, we then have have Eq.~\eqref{exponential-familiy} with $H(\bm{x},t) = U(\bm{x},t)$ to leading order.
In this case, a straightforward calculation shows that
\begin{align}
I(t) = \beta^2 \big( \av{(\partial_t U)^2} - \av{\partial_t U}^2 \big) \label{exponential-fisher},
\end{align}
In the spirit of stochastic thermodynamics \cite{Sek10,Sei12}, we write the change in the total energy of the system as
\begin{align}
d_t E(t) = d_t \av{U}_t = &\underbrace{\int d\bm{x} \ U(\bm{x},t) \partial_t P(\bm{x},t)}_{\av{\dot{\mathcal{Q}}}_t} \label{first-law} \\
& \quad + \underbrace{\int d\bm{x} \ \partial_t U(\bm{x},t) P(\bm{x},t)}_{\av{\dot{\mathcal{W}}}_t} . \n
\end{align} 
We interpret the first term (i.~e.~the change in energy due to the change in the system's state) as the rate of heat dissipated into the environment and the second term (i.~e.~the change in energy due to the driving) as the rate of work performed on the system,
\begin{align}
\dot{\mathcal{W}}(t) = \partial_t U(\bm{x}(t),t) . 
\end{align}
We then have
\begin{align}
v_I(t) = \beta \sqrt{\av{\Delta \dot{\mathcal{W}}^2}_t} \label{exponential-velocity},
\end{align}
that is, the intrinsic speed of the time evolution is set by the typical fluctuations of the input power.
At first glance, it might seem surprising that the fluctuations of the power rather than the average power determine the speed of the time evolution.
However, this is due to the close relation between averages and fluctuations for close to equilibrium systems described by the exponential family.
%We have from the fluctuation-dissipation theorem
%\begin{align}
%d_t \av{r}_t = -\beta \av{\Delta r \Delta \dot{\mathcal{W}}}_t,
%\end{align}
%which shows that it is indeed the fluctuations of the power that govern the evolution of observables.
We further note that, plugging the definition of the heat in Eq.~\eqref{first-law} into Eq.~\eqref{speed-limit-0}, we obtain in the general case,
\begin{align}
\big\vert\av{\dot{\mathcal{Q}}}_t \big\vert \leq \sqrt{\av{\Delta U^2}_t} v_I(t).
\end{align}
or, specializing to the quasistatic case Eq.~\eqref{exponential-velocity},
\begin{align}
\big\vert\av{\dot{\mathcal{Q}}}_t \big\vert \leq \beta \sqrt{\av{\Delta U^2}_t \av{\Delta \dot{\mathcal{W}}^2}_t} ,
\end{align}
i.~e.~the rate of heat exchange is bounded by the product of the fluctuations of the total energy and input power.
As a concrete application of the relation \eqref{exponential-velocity} between the intrinsic velocity and the power fluctuations, we consider the case where the only dependence on time of the potential is a slowly varying, spatially homogeneous force $f_i(t)$ on a \enquote{probe particle} $i$, i.~e.~$H(\bm{x},t) = H_0(\bm{x}) + f_i(t) x_i$.
In this case, we have
\begin{align}
v_I(t) = \beta \big\vert \dot{f}_i(t) \big\vert \sqrt{\av{\Delta x_i^2}_t} .
\end{align}
Thus, the intrinsic speed of the time evolution is set by the fluctuations of the position of the probe particle.
Using the speed limit Eq.~\eqref{speed-limit-0}, we then have for any other particle $j$,
\begin{align}
\frac{\big\vert d_t \av{x_j}_t \big\vert}{\sqrt{\av{\Delta x_j^2}_t}} \leq \beta \big\vert \dot{f}_i(t) \big\vert \sqrt{\av{\Delta x_i^2}_t} .
\end{align}
This surprising result states that, when slowly driving the probe particle, the maximal effect of this perturbation on any other particle is determined by the fluctuations of the probe particle's position, independent of the type of interactions between the particles.
In particular, by only measuring the fluctuations of the probe particle, we can estimate how strongly other particles will be affected.

For general dynamics in which the probability density does not belong to the exponential family, the relation between the Fisher information and other thermodynamic observables is not so obvious.
However, its time-derivative bears some striking resemblance to the entropy production rate.
Introducing the local mean velocity $\bm{\nu}(\bm{x},t) = \bm{\mu} (\bm{f}(\bm{x},t) - \kb T \bm{\nabla} \ln (P(\bm{x},t))$ as above, we can write $\bm{\nabla}\dot{\Phi}(\bm{x},t) = (\bm{\mu}^{-1} \dot{\bm{\nu}}(\bm{x},t) - \dot{\bm{f}}(\bm{x},t))/(\kb T)$. 
Plugging this into Eq.~\eqref{fisher-split-2}, we have
\begin{align}
\mathcal{I}(t) \equiv \frac{1}{\kb T} \Av{\dot{\bm{\nu}}^T \bm{\mu}^{-1} \dot{\bm{\nu}}}_t &= \frac{1}{\kb T} \Av{\dot{\bm{f}}^T \dot{\bm{\nu}}}_t - \frac{1}{2} d_t I(t) \label{fisher-meanvel} .
\end{align}
The expression Eq.~\eqref{fisher-meanvel} resembles the total entropy production Eq.~\eqref{entropy}, which can be written as
\begin{align}
\sigma^\text{tot}(t) = \frac{1}{\kb T} \Av{\bm{\nu}^T \bm{\mu}^{-1} \bm{\nu}}_t &= \frac{1}{\kb T} \Av{\bm{f}^T \bm{\nu}}_t + d_t S^\text{sys}(t) \label{entropy-meanvel},
\end{align}
where $S^\text{sys}(t) = - \int d\bm{x} \ \ln(P(\bm{x},t)) P(\bm{x},t)$ is the Shannon entropy with $d_t S^\text{sys}(t) = \sigma^\text{sys}(t)$.
In both cases, we have an explicitly positive quantity, which is decomposed into a total time derivative plus an additional term.
In the case of Eq.~\eqref{entropy-meanvel}, the positive quantity is the rate of total entropy production, proportional to the square of the local mean velocity.
The total time derivative is the change in Shannon entropy and the additional term can be identified with the rate of heat exchanged with the environment $\dot{\mathcal{Q}}(t) = - \av{\bm{f}^T \bm{\nu}}_t$.
For Eq.~\eqref{fisher-meanvel}, the positive quantity $\mathcal{I}(t)$ is proportional to the square of the \emph{change} in local mean velocity and the total time derivative is the change in Fisher information.
We conclude that, in contrast to the entropy production, which describes the local flows in the system, the Fisher information describes how these flows change in time.
Thus, while the total entropy production is determined by the magnitude of the local mean \emph{velocity} $\bm{\nu}$, the positive relaxational contribution to the Fisher information $\mathcal{I}(t)$ given by the magnitude of the local mean \emph{acceleration} $\dot{\bm{\nu}}$.

Integrating Eq.~\eqref{entropy-meanvel} from $t=0$ to $t=\tau$, we obtain the second law of thermodynamics from the positivity of the total entropy production,
\begin{align}
-\Delta \mathcal{Q} = \int_0^\tau dt \ \av{\bm{f}^T \bm{\nu}}_t \geq -\kb T \Delta S^\text{sys}. \label{second-law}
\end{align}
In particular, if the state of the system is the same at $t=0$ and $t=\tau$, (for example for periodically driven systems), then the Shannon entropy does not change and heat is dissipated into the environment, $\Delta \mathcal{Q} \leq 0$.
Likewise, since the left-hand side of Eq.~\eqref{fisher-meanvel} is positive, we find,
\begin{align}
\int_0^\tau dt \ \Av{\dot{\bm{f}}^T \dot{\bm{\nu}}}_t \geq \frac{\kb T}{2} \big( I(\tau) - I(0) \big) .
\end{align}
If the Fisher information is the same in the initial and final state, $\Delta I = 0$, then the product of the time derivatives of the force and the local mean velocity has to be positive on average,
\begin{align}
\int_0^\tau dt \ \Av{\dot{\bm{f}}^T \dot{\bm{\nu}}}_t \geq 0 \label{force-meanvel} .
\end{align}
This is in particular true for any process which connects two arbitrary steady states, since then we have $I(0) = I(\tau) = 0$.
In terms of the local mean velocity, we can interpret Eq.~\eqref{second-law} as stating that, on average, the local mean velocity should be parallel to the external force; Eq.~\eqref{force-meanvel} demands that the same is true for the time-derivatives of the respective quantities.

One of the ambiguities of stochastic thermodynamics is that the stochastic definition of heat and work is not unique in the sense that only their averages are constrained by thermodynamics.
This allows for different definitions of e.~g.~heat which all have the same average yet differ in terms of their fluctuations.
For example, using the local mean velocity defined above, we may define a stochastic heat as
\begin{align}
\dot{\mathcal{Q}}^{(1)}(t) = - \bm{\nu}(\bm{x}(t),t)^T \bm{\mu}^{-1} \bm{\nu}(\bm{x}(t),t) + \kb T \bm{\nabla} \cdot \bm{\nu}(\bm{x}(t),t) .
\end{align}
Taking the average, we obtain after some calculus
\begin{align}
\av{\dot{\mathcal{Q}}^{(1)}}_t = - \kb T \big( \sigma^\text{tot}(t) - \sigma^\text{sys}(t) \big),
\end{align}
where $\sigma^\text{sys}(t) = - \partial_t \int d\bm{x} \ln(P(\bm{x},t)) P(\bm{x},t)$ is the rate of Shannon entropy production.
This definition corresponds to interpreting heat as the entropy increase of the environment.
On the other hand, we may also define
\begin{align}
\dot{\mathcal{Q}}^{(2)}(t) = -\bm{f}(\bm{x}(t),t) \cdot \bm{\nu}(\bm{x}(t),t).
\end{align}
This definition corresponds to interpreting heat as the work performed by the force $\bm{f}$ against the flows in the system.
Importantly, both definitions yield the same average $\av{\dot{\mathcal{Q}}^{(1)}}_t = \av{\dot{\mathcal{Q}}^{(2)}}_t$ and are thus consistent with macroscopic thermodynamics.
However, their fluctuations differ and, somewhat surprisingly, this difference can be related to the Fisher information (see Ref.~\cite{Ito17} and Appendix \ref{app-cost-fp}),
\begin{align}
I(t) = \beta^2 \Av{\big(\dot{\mathcal{Q}}^{(1)} - \dot{\mathcal{Q}}^{(2)}\big)^2}_t .
\end{align}
Similar to Eq.~\eqref{exponential-fisher}, the Fisher information can be expressed as power fluctuations, however, in general, this power is not simply the input power but rather the difference between the \enquote{entropic} and \enquote{operational} definitions of the heat rate.

Finally, we note that the expression for $d_t I_\text{rel}(t)$ in Eq.~\eqref{fisher-split-2} is precisely the same as the factor relating the entropy production rate to the Fisher information in Eq.~\eqref{entropy-fisher}.
We thus have the relation between entropy production and Fisher information for Fokker-Planck dynamics
\begin{align}
d_t I_\text{rel}(t) \leq - \frac{2 I(t)^2}{\sigma^\text{tot}(t)} .
\end{align}
For relaxation processes in particular we have $d_t I(t) = d_t I_\text{rel}(t)$ and thus
\begin{align}
d_t I(t) \leq - \frac{2 I(t)^2}{\sigma^\text{tot}(t)} \label{fisher-rate-bound}.
\end{align}
This relation is rather surprising, since the Fisher information is defined only in terms of the probability density, while the entropy production rate depends on the explicit dynamics, i.~e.~the forces acting on the system.
Nevertheless, the two quantities are related for relaxation processes: The Fisher information has to decay at a minimum rate which is given by the ratio of the Fisher information and the total entropy production rate.

Generally, measuring the Fisher information requires determining the probability density $P(\bm{x},t)$ and its time-derivative, which may be challenging in experimental situations.
However, we can invert the speed limit Eq.~\eqref{speed-limit-0} to give a lower estimate on the Fisher information.
In particular, from the generalized Cram{\'e}r-Rao bound for several observables \cite{Kay93}, we obtain the inequality
\begin{align}
\big( d_t \av{\bm{r}}_t \big)^T \bm{\Xi}^{-1}_r(t) \big( d_t \av{\bm{r}}_t \big) \leq I(t) .
\end{align}
Here $\bm{r}(\bm{x}) = (r_1(\bm{x}), \ldots, r_K(\bm{x}))$ is a vector composed of $K$ observables and $\bm{\Xi}_r(t)$ is their covariance matrix with entries $(\bm{\Xi}_r(t))_{i j} = \av{\Delta r_i \Delta r_j}_t$.
By measuring more observables, we can thus obtain a tighter lower bound on the Fisher information.
Consequently, even if the Fisher information is not directly accessible via measurements, it nevertheless limits how much information can be obtained by measuring the time evolution of arbitrary observables.

\section{Geometric interpretation} \label{sec-fisher-def}

\begin{figure}
\includegraphics[width=.47\textwidth]{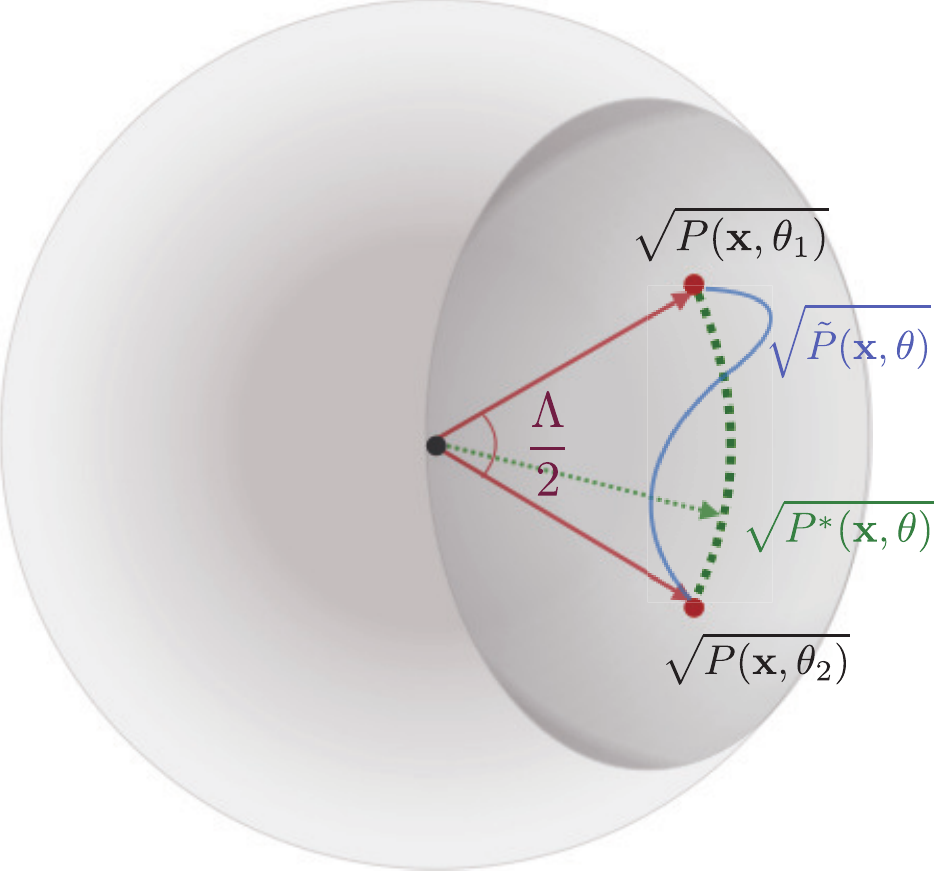}
\caption{Illustration of different parameterizations of the path between two probability densities $P(\theta_1)$ and $P(\theta_2)$ (red dots). While any parameterization $\tilde{P}$ is constrained to lie on the unit sphere due to normalization, the length of the path can be arbitrarily long (blue). By contrast, the shortest possible path is given by the geodesic $P^*$ (green dashed). Note that this three-dimensional illustration corresponds to the case of three discrete states, whereas in the case of continuous random variables the underlying space is infinite-dimensional. \label{fig-geo}}
\end{figure}

Since the Fisher information is a central quantity in information geometry \cite{Ama07}, it is natural to consider the geometric interpretation of the previous results.
We start by reviewing some general properties and the geometric interpretation of Fisher information.
Consider a system described by a probability density $P(\bm{x},\theta)$, $\theta$ is a parameter.
If $\theta$ is equal to the observation time $t \in [0,\mathcal{T}]$ then $P(\bm{x},t)$ describes the time evolution of the probability density.
However, $\theta$ may also be some other, more general parameter, e.~g.~$P(\bm{x},\theta)$ could be the steady state probability density of the system and $\theta$ some externally tunable field.
In the following, we assume that $P(\bm{x},\theta)$ depends smoothly on $\theta$, such that, in particular, the derivative $\partial_\theta P(\bm{x},\theta)$ exists and is a continuous function and the second derivative $\partial_\theta^2 P(\bm{x},\theta)$ exists.
The Fisher information $I(\theta)$ is defined by \cite{Kay93}
\begin{align}
I(\theta) &= \int d\bm{x} \ \frac{\big(\partial_\theta P(\bm{x},\theta) \big)^2}{P(\bm{x},\theta)} \label{fisher-info} \\
& = \Av{\big(\partial_\theta \ln P\big)^2}_\theta = - \Av{\vphantom{\big)^2}\partial_\theta^2 \ln P}_\theta, \n
\end{align}
where $\av{\ldots}_\theta$ denotes an average with respect to $P(\bm{x},\theta)$.
%Here, and in the following, we use the shorthand $\partial_\theta f = \partial f/\partial \theta$ for partial and $d_\theta f = d f/d \theta$ for total derivatives.
The last equality follows from the normalization of the probability density $\partial_\theta \int d\bm{x} \ P(\bm{x},\theta) = \partial_\theta 1 = 0$.
We note that, by definition, the Fisher information is positive and vanishes only if the probability density is independent of $\theta$.
The Fisher information is related to the Kullback-Leibler divergence or relative entropy between two distributions $P(\bm{x})$ and $Q(\bm{x})$,
\begin{align}
D_\text{KL}(Q \Vert P) = \int d\bm{x} \ Q(\bm{x}) \ln \bigg(\frac{Q(\bm{x})}{P(\bm{x})} \bigg) .
\end{align} 
Choosing $Q(\bm{x}) = P(\bm{x},\theta+d\theta)$, i.~e.~the probability distribution at an infinitesimally different value of $\theta$, the corresponding Kullback-Leibler divergence is to leading order in $d\theta$ given by
\begin{align}
D_\text{KL} (P(\theta + d\theta) \Vert P(\theta)) = \frac{1}{2} I(\theta) d\theta^2 + O(d\theta^3) ,
\end{align}
and the Fisher information thus is the curvature of the Kullback-Leibler divergence.
Similar to the Kullback-Leibler divergence, the Fisher information is additive in the following sense:
Suppose that we subdivide the random variables into two sets $\bm{x} = (\bm{y},\bm{\psi})$.
Introducing the conditional probability density $P_{\psi\vert y}(\bm{\psi}, \theta \vert \bm{y})$, we can then write
\begin{align}
P(\bm{x},\theta) = P_{\psi \vert y}(\bm{\psi}, \theta \vert \bm{y}) P_y(\bm{y},\theta),
\end{align}
where $P_y(\bm{y},\theta))$ is the marginal density of the random variables $\bm{y}$.
Then a straightforward calculation shows that
\begin{align}
I(\theta) = I_{\psi \vert y}&(\theta) + I_y(\theta) \quad \text{with} \label{fisher-add} \\
I_{\psi \vert y}(\theta) &\equiv \int d\bm{\psi} \int d\bm{y} \ \frac{\big(\partial_{\theta} P_{\psi \vert y}(\bm{\psi},\theta \vert \bm{y}) \big)^2}{P_{\psi \vert y}(\bm{\psi},\theta \vert \bm{y})} P_y(\bm{y},\theta) \nn
&= \Av{(\partial_\theta \ln P_{\psi \vert y} )^2}_\theta \nn
I_y(\theta) &\equiv \int d\bm{y} \ \frac{\big(\partial_{\theta} P_{y}(\bm{y},\theta) \big)^2}{P_{y}(\bm{y},\theta)} = \Av{(\partial_\theta \ln P_{y} )^2}_\theta \n .
\end{align}
The Fisher information can thus be decomposed into two positive terms, depending on the conditioned statistics of the random variables $\bm{\psi}$ and the statistics of the random variables $\bm{y}$, respectively.
In particular, we have $I(\theta) \geq I_y(\theta)$, i.~e.~eliminating variables decreases the Fisher information.
If the random variables $\bm{\psi}$ and $\bm{y}$ are further independent, then we have $I(\theta) = I_{\psi}(\theta) + I_y(\theta)$.
The geometric interpretation of the Fisher information follows from defining a statistical line element $ds$ by
\begin{align}
ds^2 = I(\theta) d\theta^2 \label{line-element}.
\end{align}
The quantity $ds$ may be thought of as a dimensionless distance between the probability densities at two infinitesimally different values of $\theta$, i.~e.~between $P(\bm{x},\theta)$  and $P(\bm{x},\theta + d\theta)$.
The infinitesimal statistical line element in a natural way defines a statistical length,
\begin{align}
\mathcal{L}(\theta_2,\theta_1) = \int_{\theta_1}^{\theta_2} ds = \int_{\theta_1}^{\theta_2} \vert d\theta \vert \ \sqrt{I(\theta)} \label{length} .
\end{align}
This length measures the length of the path traced by the probability density under a change of the parameter from $\theta = \theta_1$ to $\theta = \theta_2$.
We remark that the statistical length has all the properties expected of a path length, in that it satisfies the triangle inequality and is invariant under monotonic reparameterizations of the path.
We further remark that the above notions can be extended to a higher-dimensional parameter space; however, in what follows, we will take $\theta$ to be the evolution time of a stochastic system and thus will only require the one-dimensional case.
In principle, there are infinitely many possible parameterizations of the path from $\theta_1$ to $\theta_2$, e.~g.~$\tilde{P}(\bm{x},\theta)$ with $\tilde{P}(\bm{x},\theta_1) = P(\bm{x},\theta_1)$ and $\tilde{P}(\bm{x},\theta_2) = P(\bm{x},\theta_2)$ but $\tilde{P}(\bm{x},\theta) \neq P(\bm{x},\theta)$ otherwise.
However, since any parameterization has to give a normalized probability density, $\int d\bm{x} \ P(\bm{x},\theta) = 1$, there exists a unique parameterization that minimizes the path length $\mathcal{L}(\theta_2,\theta_1)$ (see Appendix \ref{app-minprob}).
Geometrically, the normalization condition means that $\sqrt{P(\bm{x},\theta)}$ has to be a vector of length $1$, i.~e.~tracing a path on the infinite-dimensional unit sphere, see the illustration in Fig.~\ref{fig-geo}.
Thus the minimal length is the arc length between the point $P(\bm{x},\theta_1)$ and $P(\bm{x},\theta_2)$,
\begin{align}
\Lambda \equiv  2 \arccos \bigg( \int d\bm{x} \ \sqrt{P(\bm{x},\theta_2) P(\bm{x},\theta_1)} \bigg) \label{angle} ,
\end{align}
which is also referred to as the Bhattacharyya angle \cite{Bha43}.
The parameterization that realizes this minimal length is the geodesic curve,
\begin{align}
P^*(\bm{x},\theta) &= \frac{1}{\sin\big(\frac{\Lambda}{2}\big)^2} \bigg( \sin\Big(\frac{\Lambda}{2}\frac{\theta_2-\theta}{\theta_2 - \theta_1}\Big) \sqrt{P(\bm{x},\theta_1)} \nn
& \hspace{1.5cm} + \sin\Big(\frac{\Lambda}{2} \frac{\theta-\theta_1}{\theta_2 - \theta_1}\Big) \sqrt{P(\bm{x},\theta_2)} \bigg)^2 \label{optimal-prob} ,
\end{align}
which simultaneously minimizes the action integral
\begin{align}
\mathcal{C}(\theta_2,\theta_1) = \frac{1}{2} \int_{\theta_1}^{\theta_2} d\theta \ I(\theta) \label{action}.
\end{align}
For the geodesic curve, we thus have
\begin{align}
\mathcal{L}^*(\theta_2,\theta_1) = \Lambda, \qquad \mathcal{C}^*(\theta_2,\theta_1) = \frac{\Lambda^2}{2 (\theta_2 - \theta_1)} \label{length-minimal},
\end{align}
while for any other parameterization $P(\bm{x},\theta)$, we have the inequalities
\begin{align}
\mathcal{C}(\theta_2,\theta_1) \geq \frac{\mathcal{L}^2}{2 (\theta_2 - \theta_1)} \geq \frac{\Lambda^2}{2 (\theta_2 - \theta_1)} \label{length-ineq} ,
\end{align}
where the first inequality follows from applying the Cauchy-Schwartz inequality to $\mathcal{L}^2$ and the second one is a consequence of $\mathcal{L} \geq \Lambda$.

\begin{figure}
    \centering
    \includegraphics[width=.47\textwidth]{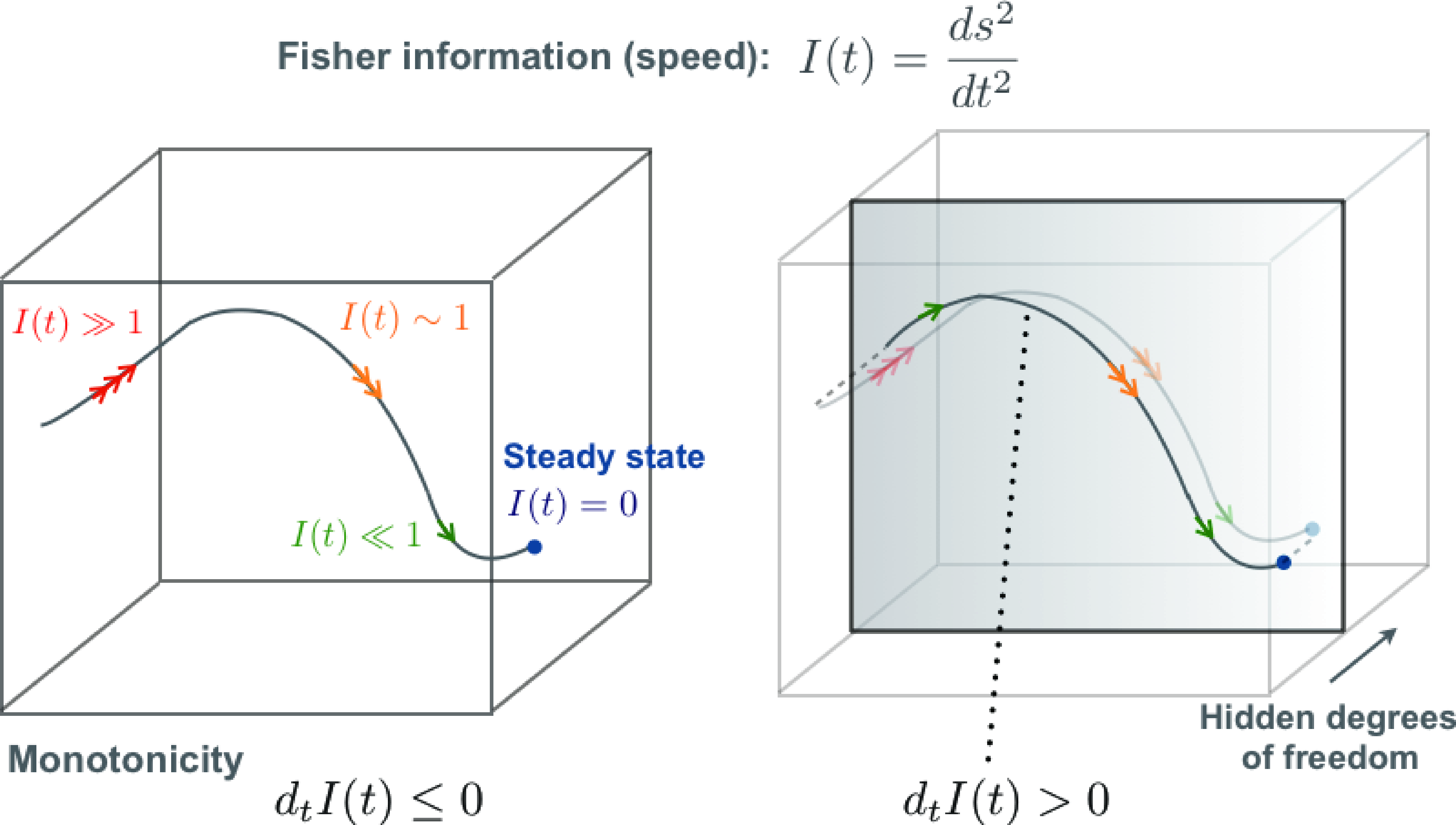}
    \caption{Illustration of the non-monotonic behavior of the Fisher information in the presence of hidden degrees of freedom. 
    The relaxation of the system towards the steady state traces a path in the (in this case three-dimensional) state space. In the full state space (left), this relaxation is accompanied by a monotonic decrease of the temporal Fisher information (i.~e.~the speed of the relaxation process), as indicated by the arrows along the path.
    Tracing out the hidden degrees of freedom corresponds to a projection of the path into a lower-dimensional subspace (shaded, right). 
    In this example, the initial relaxation process is mostly in the direction of the hidden degree of freedom, resulting in an apparently slower speed in the projected dynamics and thus in a non-monotonic behavior of the Fisher information.}
    \label{fig-hidden-geo}
\end{figure}

Applying the above to the case where $\theta = t$, we immediately have the statistical length of the time evolution of the probability density,
\begin{align}
\mathcal{L}(t) = \int_0^t ds \ v_I(s) \quad \Leftrightarrow \quad v_I(t) = d_t \mathcal{L}(t) .
\end{align}
Thus the intrinsic velocity $v_I(t)$ is the velocity of the probability density vector and measures how fast the system traverses the path.
Using the speed limit Eq.~\eqref{speed-limit-0}, we find that
\begin{align}
\mathcal{L}(t) \geq \mathcal{L}_r(t) \equiv \int_0^t ds \ \frac{\big\vert d_s \av{r}_s \big\vert}{\sqrt{\av{\Delta r^2}_s}} \label{length-bound} .
\end{align}
We can interpret $\mathcal{L}_r$ as the \enquote{length} of the time evolution of the observable $r$.
In Ref.~\cite{Cro07}, a protocol to measure the statistical length for probability distributions belonging to the exponential family Eq.~ \eqref{exponential-familiy} was suggested.
While the above only provides a lower estimate, we can obtain it for arbitrary dynamics and any observable.
On the other hand, this also provides a geometric interpretation of the speed limit Eq.~\eqref{speed-limit-0}.
The statistical length $\mathcal{L}$ is the length of the path traced by the probability in a high-dimensional space.
Since computing the average corresponds to a projection into a lower-dimensional space (in this case one-dimensional), Eq.~\eqref{speed-limit-0} states that the projected path $\mathcal{L}_R$ is always shorter.
In a similar way, the non-monotonic behavior in the presence of hidden degrees of freedom discussed in Section \ref{sec-monotonicity} can be understood from a geometric point of view, as illustrated in Fig.~\ref{fig-hidden-geo}.
Moreover, if the observable $r$ only depends on a subset $\bm{y}$ of the stochastic variables, then we have the sequence of bounds
\begin{align}
\frac{\big\vert d_t \av{r}_t \big\vert}{\sqrt{\av{\Delta r^2}_t}} \equiv v_r(t) \leq v_{I_r}(t) \leq v_{I_y}(t) \leq v_{I}(t) .
\end{align}
Here $v_{I_r}$ is the evolution speed of $P_r(r,t) = \int d\bm{x} \ \delta(r(\bm{x}) - r) P(\bm{x},t)$ and $v_{I_y}$ is the evolution speed of $P_y(\bm{y},t)$.
This means that as less information is included in the probability density, its evolution speed gets slower.
This is intuitive as averaging out degrees of freedom may eliminate fast timescales from the dynamics.

For $\theta = t$, the action functional Eq.~\eqref{action} plays the role of a kinetic energy,
\begin{align}
\mathcal{C}(t) = \frac{1}{2} \int_0^t ds \ v_I(s)^2 ,
\end{align}
and from Eq.~\eqref{length-ineq}, we immediately get the integral speed limit derived in Ref.~\cite{Ito17},
\begin{align}
t \geq \frac{\mathcal{L}^2}{2 \mathcal{C}} \geq \frac{\Lambda^2}{2 \mathcal{C}} \label{speed-limit-int},
\end{align}
which provides a lower bound on the time it takes for the probability vector to trace a path of length $\mathcal{L}$.
Using Eq.~\eqref{length-ineq}, we can further derive an integral speed limit in terms of the observable $r$,
\begin{align}
t \geq \frac{\mathcal{L}_r^2}{2 \mathcal{C}} \geq \frac{1}{2 \mathcal{C}} \frac{\big(\av{r}_t - \av{r}_0\big)^2}{\av{\Delta r^2}_\text{max}},
\end{align}
where $\av{\Delta r^2}_\text{max}$ denotes the maximal variance of $r$ in the interval $[0,t]$.
This provides a lower bound on the time necessary to change the average of the observable from $\av{r}_0$ to $\av{r}_t$.
In case of a relaxation process, we can combine Eqs.~\eqref{fisher-monotonic} and \eqref{speed-limit-int} in order to obtain,
\begin{align}
t &\geq \frac{\Lambda^2}{2 \mathcal{C}} \geq \frac{\Lambda^2}{t I(0)} \nn
\Rightarrow t &\geq \frac{\Lambda}{v_I(0)} \label{speed-limit-initial} .
\end{align}
Since the Bhattacharyya angle Eq.~\eqref{angle} only depends on the initial and final state, Eq.~\eqref{speed-limit-initial} constitutes a speed limit on relaxation processes, which only depends on the initial speed of the evolution.
We remark that such speed limits have been extensively discussed in quantum-mechanical systems (see e.~g.~Ref.~\cite{Pri16}), however, it has recently been found that similar bounds also apply to classical and stochastic dynamics \cite{Sha18,Oku18}.
In contrast to the Margolus-Levitin-type bound derived in Ref.~\cite{Oku18} (Eq.~(23) therein), this result does not require any particular spectral properties of the generator or existence of a steady state; the only requirement is that the generator does not depend explicitly on time.
Note that, in contrast to a jump process on a finite state space, in the continuous case, the existence of a steady state is not guaranteed even if the generator is time-independent and ergodic; the simplest example is Brownian motion in an infinite domain, which does not possess a steady state probability density.
The bound \eqref{speed-limit-initial} is further tighter than the Mandelstam-Tamm-type bound derived in Ref.~\cite{Oku18} (Eq.~(26) therein) for a particle relaxing in a binding potential, since we have $2\arccos(x) \geq \pi(1-x)$ for $x > 0$.
Using Eq.~\eqref{fisher-rate-bound}, we further have that
\begin{align}
\bigg( \int_0^t dt' \ I(t') \bigg)^2 &\leq \bigg( \int_0^t dt' \ \sqrt{-\frac{1}{2} d_{t'} I(t') \sigma^\text{tot}(t')}  \bigg)^2 \nn
&\leq -\frac{1}{2} \int_0^t dt' \ d_{t'} I(t') \int_0^t dt' \ \sigma^\text{tot}(t') \nn
&= \frac{1}{2} \big( I(0) - I(t) \big) \Delta S^\text{tot} \nn
&\leq \frac{1}{2} I(0) \Delta S^\text{tot},
\end{align}
where we used the Cauchy-Schwarz inequality from the first to the second line.
From Eq.~\eqref{speed-limit-int} this gives us
\begin{align}
t \geq \frac{\sqrt{2}\Lambda^2}{\sqrt{I(0) \Delta S^\text{tot}}},
\end{align}
or combining with Eq.~\eqref{speed-limit-initial},
\begin{align}
t \geq \frac{\Lambda}{v_I(0)} \times \text{max} \bigg(1, \sqrt{ \frac{2 \Lambda^2}{\Delta S^\text{tot}} } \bigg) .
\end{align}
Thus, we can possibly obtain a tighter speed limit by using additional information on the thermodynamic properties of the relaxation process in the form of the total entropy production.

\section{Examples} \label{sec-examples}

\subsection{General normal distributions}
A particularly succinct and widely applicable example for the relation between statistical length, Fisher information and observables is for a normal distribution in $M$ variables,
\begin{align}
P(\bm{x},t) &= \frac{1}{\sqrt{(2 \pi)^M \det(\bm{\Xi}(t))}} \label{normal-dist} \\
&\qquad \times \exp \bigg[-\frac{1}{2} \big(\bm{x} - \bm{\mu}(t)\big)^T \bm{\Xi}(t)^{-1} \big(\bm{x} - \bm{\mu}(t)\big) \bigg] \n
\end{align}
with the average $\av{\bm{x}}_t = \bm{\mu}(t)$ and the (symmetric and positive definite) covariance matrix $\bm{\Xi}(t)$ defined by
\begin{align}
\Xi_{i j}(t) = \Av{\big(x_i - \mu_i(t) \big) \big(x_j - \mu_j(t) \big)}_t.
\end{align}
Here the subscript $T$ denotes transposition and $\det$ the determinant.
In this case, we can compute the rate of Shannon entropy change $\sigma^\text{sys}(t) = d_t \Sigma^\text{sys}(t)$ and the Fisher information explicitly \cite{Mal15},
\begin{subequations}
\begin{align} 
&\sigma^\text{sys}(t) = \frac{1}{2} d_t \ln\big( \det(\bm{\Xi}) \big) = \frac{1}{2} \text{tr}\big(\bm{\Xi}(t)^{-1} \dot{\bm{\Xi}}(t) \big) \label{shannon-normal}  \\
&I(t) = \dot{\bm{\mu}}(t)^T \bm{\Xi}(t)^{-1} \dot{\bm{\mu}}(t) + \frac{1}{2} \text{tr}\big( \bm{\Xi}^{-1}(t) \dot{\bm{\Xi}}(t) \bm{\Xi}^{-1}(t) \dot{\bm{\Xi}}(t)  \big) \label{fisher-normal} ,
\end{align}%
\end{subequations}
where $\dot{\bm{\mu}}(t)$ and $\dot{\bm{\Xi}}(t)$ are the component-wise time derivatives of the respective quantities and $\text{tr}$ is the trace.
A normal distribution can arise from the solution of a Fokker-Planck equation with linear drift coefficients
\begin{align}
\partial_t P(\bm{x},t) &= - \partial_{x_i} \Big( a_i(\bm{x},t) - B_{i j}(t) \partial_{x_j} \Big) P(\bm{x},t) \\
\text{with} \quad &a_i(\bm{x},t) = K_{i j}(t) x_j + k_i(t) \n 
\end{align}
with a symmetric, positive semidefinite matrix $\bm{B}$, provided that the initial distribution is normal
\begin{align}
P_0(\bm{x}) &= \frac{1}{\sqrt{(2 \pi)^M \det(\bm{\Xi}_0)}} \\
&\qquad \times \exp \bigg[-\frac{1}{2} \big(\bm{x} - \bm{\mu}_0\big)^T \bm{\Xi}_0^{-1} \big(\bm{x} - \bm{\mu}_0\big) \bigg] \n .
\end{align}
The mean and covariance matrix then are determined by the differential equations
\begin{subequations}
\begin{align}
d_t \mu_i(t) &= K_{i j}(t) \mu_j(t) + k_i(t)\\
 d_t \Xi_{i j}(t) &= K_{i l}(t) \Xi_{l j}(t) + K_{j l}(t) \Xi_{l i}(t) \\
 &\qquad + \big( B_{i j}(t) + B_{j i}(t) \big) \n ,
\end{align}
\end{subequations}
or in matrix notation (using that $\bm{B}$ is symmetric)
\begin{subequations}
\begin{align}
\dot{\bm{\mu}}(t) &= \bm{K}(t) \bm{\mu}(t) + \bm{k}(t)\\
 \dot{\bm{\Xi}}(t) &= \bm{K}(t) \bm{\Xi}(t) + \bm{\Xi}(t) \bm{K}^T(t) + 2\bm{B}(t),
\end{align}
\end{subequations}
with initial condition $\bm{\mu}(0) = \bm{\mu}_0$ and $\bm{\Xi}(0) = \bm{\Xi}_0$.
These equations allow us to write the Fisher information without relying on time-derivatives,
\begin{align}
I(t) &= \big(\bm{K} \bm{\mu} + \bm{k}\big)^T \bm{\Xi}^{-1} \big(\bm{K} \bm{\mu} + \bm{k}\big)  \\ 
& \, + \frac{1}{2} \text{tr} \Big[ \big( \bm{\Xi}^{-1} \bm{K} \bm{\Xi} + \bm{K}^T \big)^2  \nn
& \qquad + 4 \bm{B} \big( \bm{\Xi}^{-1} \bm{K} + \bm{K}^T \bm{\Xi}^{-1} \big) + 4\bm{B} \bm{\Xi}^{-1} \bm{B} \bm{\Xi}^{-1} \Big] \n .
\end{align}
Obviously, any normal distribution is uniquely determined by its mean and covariance matrix and thus the latter two quantities also specify the average of any observable $\bm{R}(\bm{x})$ and its time evolution.
However, how precisely the time evolution of the mean and covariance matrix impact the time evolution of $\av{\bm{R}}_t$, i.~e.~the explicit expression of $\av{\bm{R}}_t$ in terms of $\bm{\mu}$ and $\bm{\Xi}$ is not obvious except in simple cases.
Nevertheless, from Eq.~\eqref{speed-limit-0}, we always have the bound
\begin{align}
\av{\dot{\bm{R}}}^T \bm{\Xi}_R^{-1} \av{\dot{\bm{R}}} \leq \dot{\bm{\mu}}^T \bm{\Xi}^{-1} \dot{\bm{\mu}} + \frac{1}{2} \text{tr}\big( \bm{\Xi}^{-1} \dot{\bm{\Xi}} \bm{\Xi}^{-1} \dot{\bm{\Xi}}  \big) \label{average-bound} .
\end{align}
This bound is particularly instructive for a time-independent covariance matrix $\dot{\bm{\Xi}} = 0$, where it states that the change in the average of any observable, relative to its covariance matrix, is always less than the respective quantity for the mean of the distribution.
In this sense, no observable can change faster than the mean of the distribution.
We further note a result valid for any probability distribution which depends on time only via its mean $\bm{\mu}$, and can thus be written as $P(\bm{x},t) = \tilde{P}(\bm{x}-\bm{\mu}(t))$.
For such a probability distribution, the Fisher information is always larger than for a normal distribution with the same mean and variance,
\begin{align}
I(t) \geq I_\text{normal}(t) = \dot{\bm{\mu}}(t)^T \bm{\Xi}(t)^{-1} \dot{\bm{\mu}}(t) \label{fisher-bound} .
\end{align}
Thus, a normal distribution minimizes the Fisher information for pure translations. 
We give the proof of this result in Appendix \ref{app-logvar-normal}.
Note that the inequality \eqref{fisher-bound} breaks down if the variance or some higher cumulants depend on time.

The speed limit Eq.~\eqref{speed-limit-0} also applies to the rate of change of the Shannon entropy, since $\sigma^\text{sys}(t) = d_t - \int d\bm{x} \ln (P(\bm{x},t) ) P(\bm{x},t) = - \int d\bm{x} \ \ln (P(\bm{x},t) ) \partial_t P(\bm{x},t)$,
\begin{align}
\big(\sigma^\text{sys}(t) \big)^2 \leq \bigg( \Av{(\ln P)^2}_t - \Av{\ln P}_t^2 \bigg) I(t) .
\end{align}
For a normal distribution, this relation takes a particularly simple form, since, as we show in Appendix \ref{app-logvar-normal}, we have
\begin{align}
\Av{(\ln P)^2}_t - \Av{\ln P}_t^2 = \frac{M}{2},
\end{align}
independent of the covariance matrix.
For a normal distribution, we thus have the relation between Shannon entropy and Fisher information
\begin{align}
\big(\sigma^\text{sys}(t)\big)^2 \leq \frac{M}{2} I(t) \label{shannon-bound-normal} .
\end{align}
Using $\Sigma^\text{sys}(\mathcal{T}) - \Sigma^\text{sys}(0) = \int_0^\mathcal{T} dt \ \sigma^\text{sys}(t)$ and applying the Cauchy-Schwarz inequality, this yields
\begin{align}
\mathcal{T} \geq \frac{\big(\Sigma^\text{sys}(\mathcal{T}) - \Sigma^\text{sys}(0)\big)^2}{M \mathcal{C}}
\end{align}
Since we generally expect both $\mathcal{C}$ and $\Sigma^\text{sys}$ to scale linearly with the number $M$ of degrees of freedom, we can write this in terms of the following speed limit for normal distributions
\begin{align}
\mathcal{T} \geq \frac{\big( \bar{\Sigma}^\text{sys}(\mathcal{T}) - \bar{\Sigma}^\text{sys}(0) \big)^2}{\bar{\mathcal{C}}}, 
\end{align}
where we $\bar{\Sigma}^\text{sys} = \Sigma^\text{sys}/M$ and $\bar{\mathcal{C}} = \mathcal{C}/M$ are the Shannon entropy and thermodynamic cost per degree of freedom.
This result has two interesting consequences:
First, it provides a speed limit in terms of the Shannon entropy difference between initial and final state.
Second, it explicitly demonstrates that, at least in the case of a normal distribution, this speed limit remains useful in the limit of a macroscopic number of degrees of freedom $M \gg 1$.
We stress that the latter statement is not self-evident: For the case of the speed limit Eq.~\eqref{speed-limit-initial}, the numerator is obviously bounded from above by $\pi$, the largest possible arc length on the unit sphere.
On the other hand, the denominator scales as $\sqrt{M}$ for independent degrees of freedom, since the Fisher information is additive in this case.
Thus the right-hand side of Eq.~\eqref{speed-limit-initial} is typically of order $1/\sqrt{M}$ and the bound becomes meaningless in the macroscopic limit.

\subsection{Brownian motion} \label{sec-diffusion}
The most basic example of a continuous-valued random process is Brownian motion.
Let us first consider the classical case of an overdamped particle in a environment at temperature $T$, described by the diffusion equation
\begin{align}
\partial_t P(x,t) = -\frac{1}{\gamma} \partial_x \Big( F_0 - T \partial_x \Big) P(x,t) ,
\end{align}
or, equivalently the overdamped Langevin equation
\begin{align}
\gamma \dot{x}(t) = F_0 + \sqrt{2 \gamma T} \xi(t)
\end{align}
where $\gamma$ is the friction coefficient, $F_0$ is a constant bias force, $T$ is the temperature and $\xi(t)$ is Gaussian white noise.
The solution of the diffusion equation is straightforward,
\begin{align}
P(x,t) &= \frac{1}{\sqrt{2 \pi (2 D_x t + \av{\Delta x^2}_0)}} \\
&\qquad \times \exp \lrbra{ - \frac{\Big(x-\big(\frac{F_0}{\gamma} t + \av{x}_0\big)\Big)^2}{2 \big(2 D_x t + \av{\Delta x^2}_0\big)} } \n,
\end{align}
where $\av{x}_0$ and $\av{\Delta x^2}_0$ are the initial average and variance of the particle's position at time $t = 0$.
Here, we introduced the diffusion coefficient $D_x$ given by the Einstein relation $D_x = T/\gamma$.
As we only have one degree of freedom, the expression for the Fisher information Eq.~\eqref{fisher-normal} simplifies to
\begin{align}
I(t) = \frac{F_0^2}{2 \gamma T \left( t + \frac{\av{\Delta x^2}_0}{2 D_x} \right) } + \frac{1}{2 \left( t + \frac{\av{\Delta x^2}_0}{2 D_x} \right)^2} .
\end{align}
Both with and without bias, the Fisher information for Brownian motion is a monotonously decaying function and thus (biased) Brownian motion is a generalized relaxation process.
Note that even though the Fisher information decreases, the time-derivative of the average position $d_t \av{x}_t = F_0/\gamma$ does not decay to zero but remains constant.
This is not in contradiction with the speed limit Eq.~\eqref{speed-limit-0}, which only demands that the time derivative of $\av{x}_t$ relative to the fluctuations of $x$---which in this case increase with time---should decrease along with the Fisher information.

\subsection{Particle in a parabolic trap} \label{sec-ex-harmonic}
As a second paradigmatic example, we consider a single overdamped particle with position $x(t)$ in a parabolic trap $U(x,t) = \kappa(t)(x-r(t))^2/2$,
\begin{align}
\partial_t P(x,t) = \frac{1}{\gamma} \partial_x \Big( \kappa(t) (x - r(t)) + T(t) \partial_x \Big) P(x,t),
\end{align}
or, equivalently, the Langevin equation
\begin{align}
\gamma \dot{x}(t) = - \kappa(t) (x(t)-r(t)) + \sqrt{2 \gamma T(t)} \xi(t),
\end{align}
where $\gamma$ is the friction coefficient, $\kappa$ the spring constant and $T$ the temperature.
We allow the spring constant, temperature and equilibrium position $r(t)$ of the trap to change as a function of time.
Provided that the initial state is given by a normal distribution with average $\av{x}_0$ and variance $\av{\Delta x^2}_0$, the solution to this problem is the normal distribution
\begin{align}
P(x,t) = \frac{1}{\sqrt{2 \pi \av{\Delta x^2}_t}} \exp \Bigg[{-\frac{(x - \av{x}_t)^2}{2 \av{\Delta x^2}_t}} \Bigg] ,
\end{align}
where the average and variance of the position obey the following differential equations,
\begin{subequations}
\begin{align}
d_t \av{x}_t &= - \frac{\kappa(t)}{\gamma} \big(\av{x}_t - r(t) \big) \\
 d_t \av{\Delta x^2}_t &= - \frac{2 \kappa(t)}{\gamma} \av{\Delta x^2}_t + \frac{2 T(t)}{\gamma}  .
\end{align} \label{av-var-eqs}%
\end{subequations}
Again, for a single degree of freedom, the expression for the Fisher information is immediate from Eq.~\eqref{fisher-normal}
\begin{align}
I(t) &= \frac{1}{2} \bigg(\frac{d_t \av{\Delta x^2}_t}{\av{\Delta x^2}_t} \bigg)^2 + \frac{\big(d_t \av{x}_t \big)^2}{\av{\Delta x^2}_t} %\nn
%&= \frac{1}{\gamma^2} \Bigg( 2 \bigg( \frac{T(t)}{\av{\Delta x^2}_t} - \kappa(t) \bigg)^2 + \frac{\kappa^2}{\av{\Delta x^2}_t} \Big( \av{x}_t - r(t) \Big)^2 \Bigg).
\end{align}
Here, the Fisher information (and thus the thermodynamic cost $\mathcal{C}$) consists of two positive terms: The first one is non-zero if the variance changes as a function of time, the second one if the average position changes.
The average rates of change of Shannon $\sigma^\text{sys}(t) = d_t \Sigma^\text{sys}(t)$ and total entropy $\sigma^\text{tot}(t) = d_t \Sigma^\text{tot}(t)$ (see Appendix \ref{app-cost-fp}) are given by
\begin{align}
\sigma^\text{sys}(t) &= \frac{1}{2} \frac{d_t \av{\Delta x^2}_t}{\av{\Delta x^2}_t}\\
 \sigma^\text{tot}(t) &= \frac{\gamma \av{\Delta x^2}_t}{T(t)} \Bigg( \frac{1}{4} \bigg( \frac{d_t \av{\Delta x^2}_t}{\av{\Delta x^2}_t} \bigg)^2 + \frac{\big(d_t \av{x}_t \big)^2}{\av{\Delta x^2}_t} \Bigg) \n .
\end{align}
In this case, the bound Eq.~\eqref{shannon-bound-normal} on the rate of change of the Shannon entropy is obvious, since we have ($M = 1$)
\begin{align}
I(t) &= \frac{1}{2} \bigg(\frac{d_t \av{\Delta x^2}_t}{\av{\Delta x^2}_t} \bigg)^2 + \frac{\big(d_t \av{x}_t \big)^2}{\av{\Delta x^2}_t} \\
& \geq \frac{1}{2} \bigg(\frac{d_t \av{\Delta x^2}_t}{\av{\Delta x^2}_t} \bigg)^2 = 2 \big(\sigma^\text{sys}(t)\big)^2 \n .
\end{align}
In the case of a single Gaussian degree of freedom, we thus have equality in Eq.~\eqref{shannon-bound-normal} if the average position does not change in time.
On the other hand, the local change in Shannon and total entropy, defined in Appendix \ref{app-cost-fp}, is given by
\begin{align}
\Delta \Sigma^{\text{sys}}_\text{loc}(x,t) &= \frac{x - \av{x}_t}{ \av{\Delta x^2}_t} \\
 \Delta \Sigma^{\text{tot}}_\text{loc}(x,t) &= \frac{\gamma}{T(t)} \bigg(\frac{d_t \av{\Delta x^2}_t}{2 \av{\Delta x^2}_t} \big(x - \av{x}_t \big) + d_t \av{x}_t \bigg) \n .
\end{align}
The local change in Shannon entropy vanishes only if the particle is located at the instantaneous average position, since this corresponds to the maximum of the probability distribution and thus a slight change of the particle's position will not change its Shannon entropy.
On the other hand, the local change in total entropy always vanishes independent of the particle's position if the system is in an equilibrium state $d_t \av{x}_t = d_t \av{\Delta x^2}_t = 0$.
This reflects the fact that in an equilibrium system, the total entropy production is zero not only on average but also for every single trajectory.
Using the equations of motion \eqref{av-var-eqs}, we can also write the Fisher information as
\begin{align}
I(t) &= \frac{2}{\gamma^2} \bigg( \frac{T(t)}{\av{\Delta x^2}_t} - \kappa(t) \bigg)^2 + \frac{\kappa(t)^2}{\gamma^2 \av{\Delta x^2}_t} \Big( \av{x}_t - r(t) \Big)^2 .
\end{align}
Then the time-derivative of the Fisher information can be calculated as
\begin{align}
d_t I(t) &=\frac{2 d_t \av{\Delta x^2}_t}{\gamma \av{\Delta x^2}_t^2} \dot{T}(t) - \frac{2 \kappa(t) d_t \av{x}_t}{\gamma \av{\Delta x^2}_t} \dot{r}(t)  \\
&\qquad + \Bigg(\frac{2 \big(d_t \av{x}_t \big)^2}{\kappa(t) \av{\Delta x^2}_t} -\frac{2 d_t \av{\Delta x^2}_t}{\gamma \av{\Delta x^2}_t} \Bigg) \dot{\kappa}(t) \nn
&\qquad -\frac{2 T(t)}{\gamma} \Bigg( \frac{\big(d_t \av{\Delta x^2}_t \big)^2}{\av{\Delta x^2}_t^3} + \frac{\big(d_t \av{x}_t \big)^2}{\av{\Delta x^2}_t^2} \Bigg) \n . 
%&= \frac{2 T(t)}{\gamma \av{\Delta x^2}_t} \Bigg( \frac{d_t \av{\Delta x^2}_t}{\av{\Delta x^2}_t} \frac{\dot{T}(t)}{T(t)} + \Bigg(\frac{\gamma \big(d_t \av{x}_t \big)^2}{\kappa(t) T(t)} -\frac{d_t \av{\Delta x^2}_t}{T(t)} \Bigg) \dot{\kappa}(t) - \frac{\kappa(t) d_t \av{x}_t}{\gamma \av{\Delta x^2}_t} \dot{r}(t) - \Bigg( \frac{\big(d_t \av{\Delta x^2}_t \big)^2}{\av{\Delta x^2}_t^2} + \frac{\big(d_t \av{x}_t \big)^2}{\av{\Delta x^2}_t} \Bigg),
\end{align}
The first three terms depend explicitly on the time-derivative of $T$, $r$ and $\kappa$, respectively, while the last term is negative.
In particular, if the parameters $T$, $r$ and $\kappa$ are independent of time, then we have
\begin{align}
d_t I(t) &= -\frac{2 T}{\gamma} \Bigg( \frac{\big(d_t \av{\Delta x^2}_t \big)^2}{\av{\Delta x^2}_t^3} + \frac{\big(d_t \av{x}_t \big)^2}{\av{\Delta x^2}_t^2} \Bigg) \leq 0 ,
\end{align}
and the Fisher information decreases monotonically, as predicted by Eq.~\eqref{fisher-monotonic}.

\begin{figure}
\includegraphics[width=0.47\textwidth]{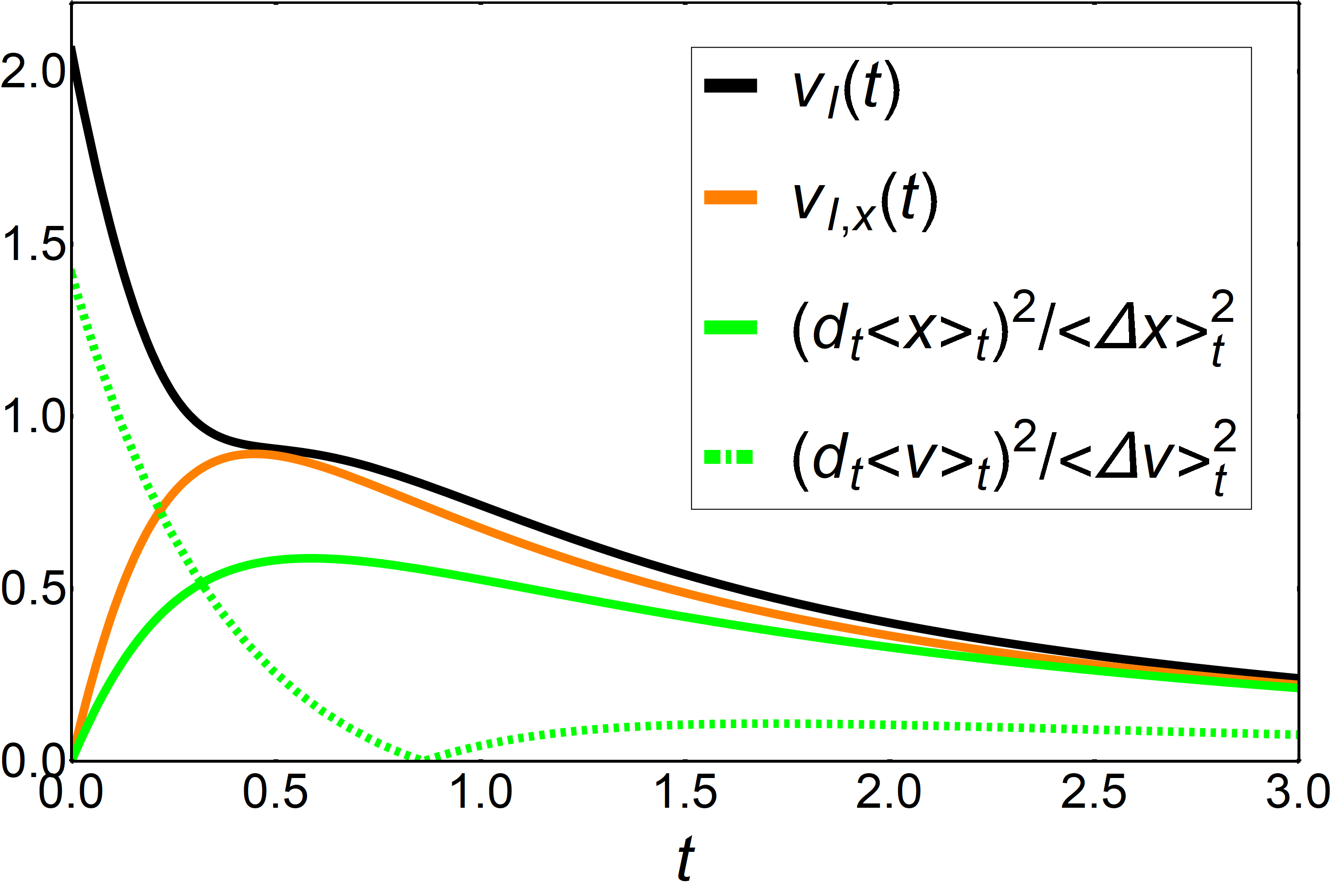}
\includegraphics[width=0.47\textwidth]{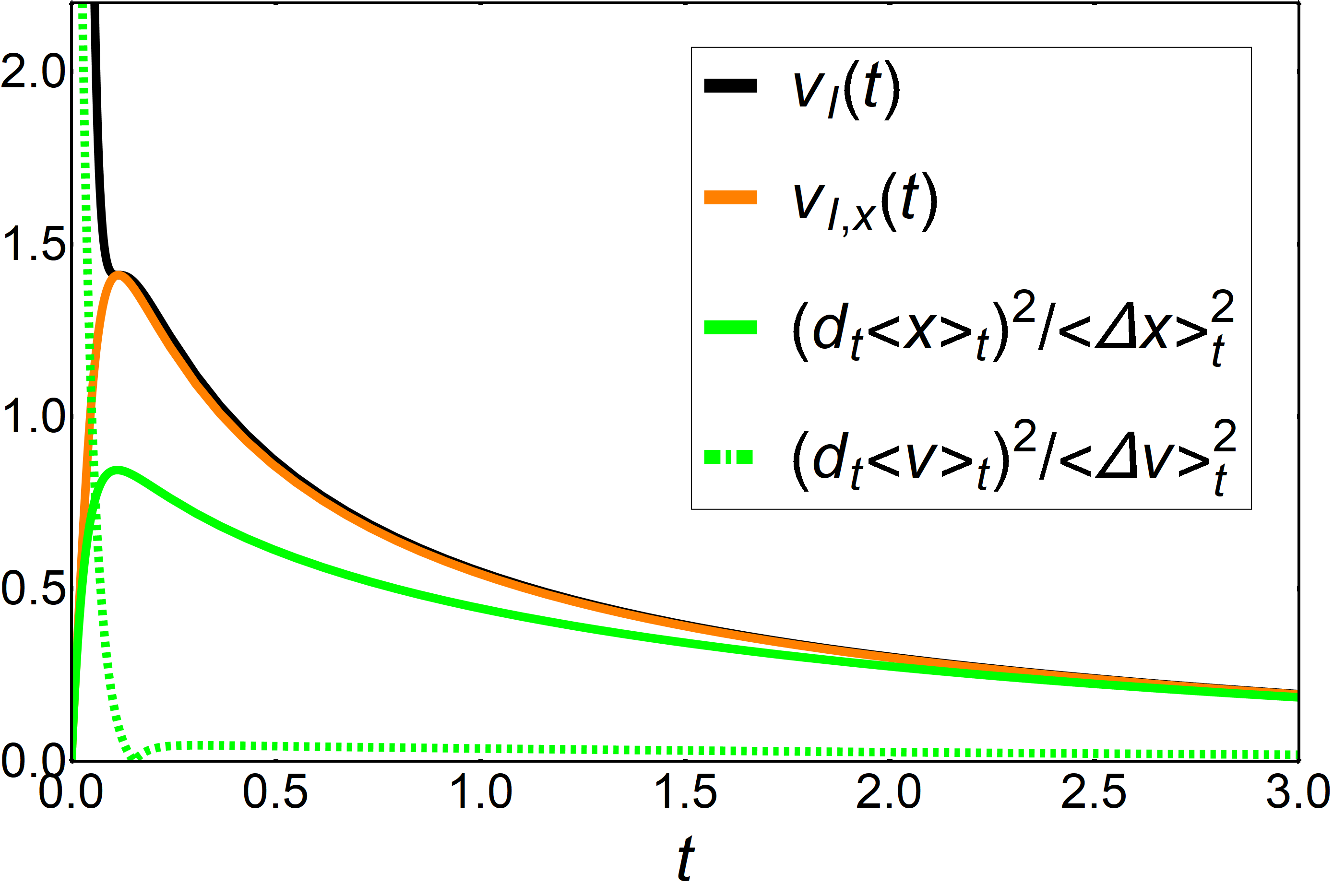}
\caption{Time-dependence of the Fisher information $I(t)$ of the joint distribution (black) and the marginal $x$-distribution (orange) for an underdamped particle in a parabolic trap for $m = 1$ (top) and $m = 1/10$ (bottom). The remaining parameters are given by $T = 2$, $\gamma = 1$, $\kappa = 1$, $r = 0$ and the system is initially in the equilibrium state corresponding to $\kappa = 4$ and $r = 2$. \label{fig-harm}}
\end{figure}
The same calculation can be done for an underdamped particle with position $x(t)$ and velocity $v(t)$
\begin{align}
\partial_t P(x,v,t) = \bigg(&- v \partial_x + \frac{\kappa(t)}{m} (x-r(t)) \partial_v \\ &
+ \frac{\gamma}{m} \partial_v \Big(  v + \frac{\gamma T(t)}{m} \partial_v \Big) \bigg) P(x,v,t) \n ,
\end{align}
with the associated equations of motion for the moments
\begin{align}
&d_t \av{x}_t = \av{v}_t, \\
&d_t \av{v}_t = - \frac{\gamma}{m} \av{v}_t - \frac{\kappa(t)}{m} \big(\av{x}_t - r(t) \big),  \nn
&d_t \av{\Delta x^2}_t = 2 \av{\Delta x \Delta v}_t, \nn
&d_t \av{\Delta x \Delta v}_t  = -\frac{\gamma}{m} \av{\Delta x \Delta v}_t - \frac{\kappa(t)}{m} \av{\Delta x^2}_t + \av{\Delta v^2}_t,  \nn
&d_t \av{\Delta v^2}_t = - \frac{2 \gamma}{m} \av{\Delta v^2}_t - \frac{2 \kappa(t)}{m} \av{\Delta x \Delta v}_t + \frac{2 \gamma T(t)}{m^2} \n .  
\end{align}
Note that the overdamped case is obtained by taking the limit of vanishing particle mass $m \rightarrow 0$.
In this case, the solution of the equations is already quite involved and we refrain from writing down the cumbersome expression for the Fisher information, which can be obtained from Eq.~\eqref{fisher-normal}.
However, in this case, since we have two degrees of freedom, already the case where $T$, $r$ and $\kappa$ do not depend on time offers some interesting insights. In this case, we observe a relaxation from the initial state to the equilibrium state with $\av{x}_\text{eq} = r$, $\av{v}_\text{eq} = 0$, $\av{\Delta x^2}_\text{eq} = T/\kappa$, $\av{\Delta x \Delta v}_\text{eq} = 0$ and $\av{\Delta v^2}_\text{eq} = T/m$.
For a non-equilibrium initial condition corresponding to potential with $\tilde{\kappa} > \kappa$ and $\tilde{r} \neq r$, the speed $v_I$ of the relaxation process is shown in Fig.~\ref{fig-harm}.
While the overall speed $v_I$ decays monotonically, as prediced by Eq.~\eqref{fisher-monotonic}, the speed $v_{I,x}$ of evolution of the marginal position distribution $P(x,t)$ exhibits a non-monotonic behavior.
Note that we also have $v_{I,x} \leq v_{I}$ i.~e.~eliminating the velocity reduces the evolution speed of the probability density.
As the system approaches the overdamped limit of vanishing mass (bottom panel) the maximum in the speed of the marginal distribution moves to shorter times and we recover the monotonic behavior of the overdamped system for times longer than the typical relaxation time of the velocity, $m/\gamma$.
As an example of the speed limit Eq.~\eqref{speed-limit-0}, we show the time-derivatives of the average position and velocity relative to their variances (green lines in Fig.~\ref{fig-harm}).
We observe that while both $x$ and $v$ obey the bound set by $v_I$, only the position $x$ obeys the tighter bound set by $v_{I,x}$.

\section{Hidden states in a molecular motor model} \label{sec-hidden-model}

\begin{figure*}
\includegraphics[width=0.47\textwidth]{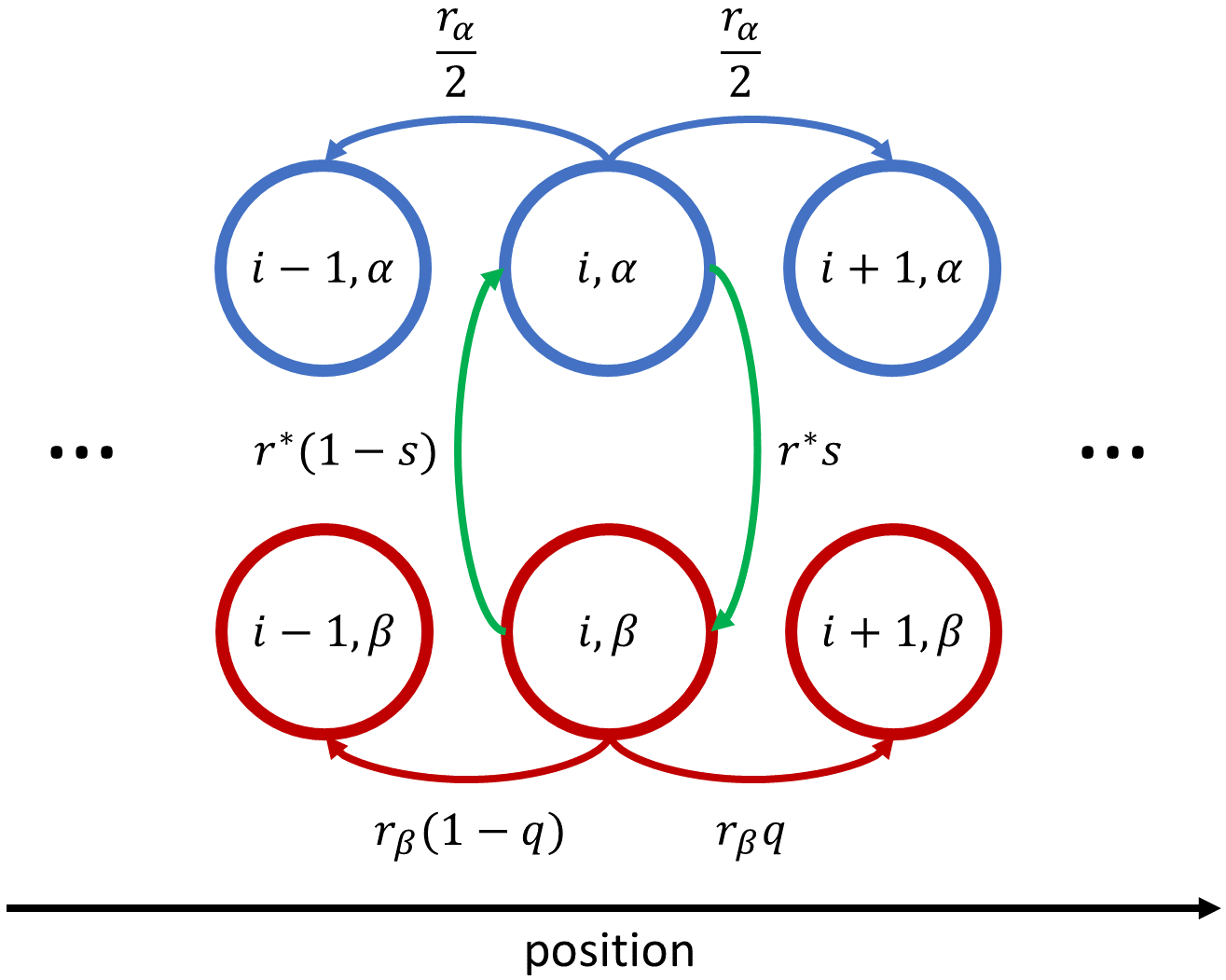}
\includegraphics[width=0.47\textwidth]{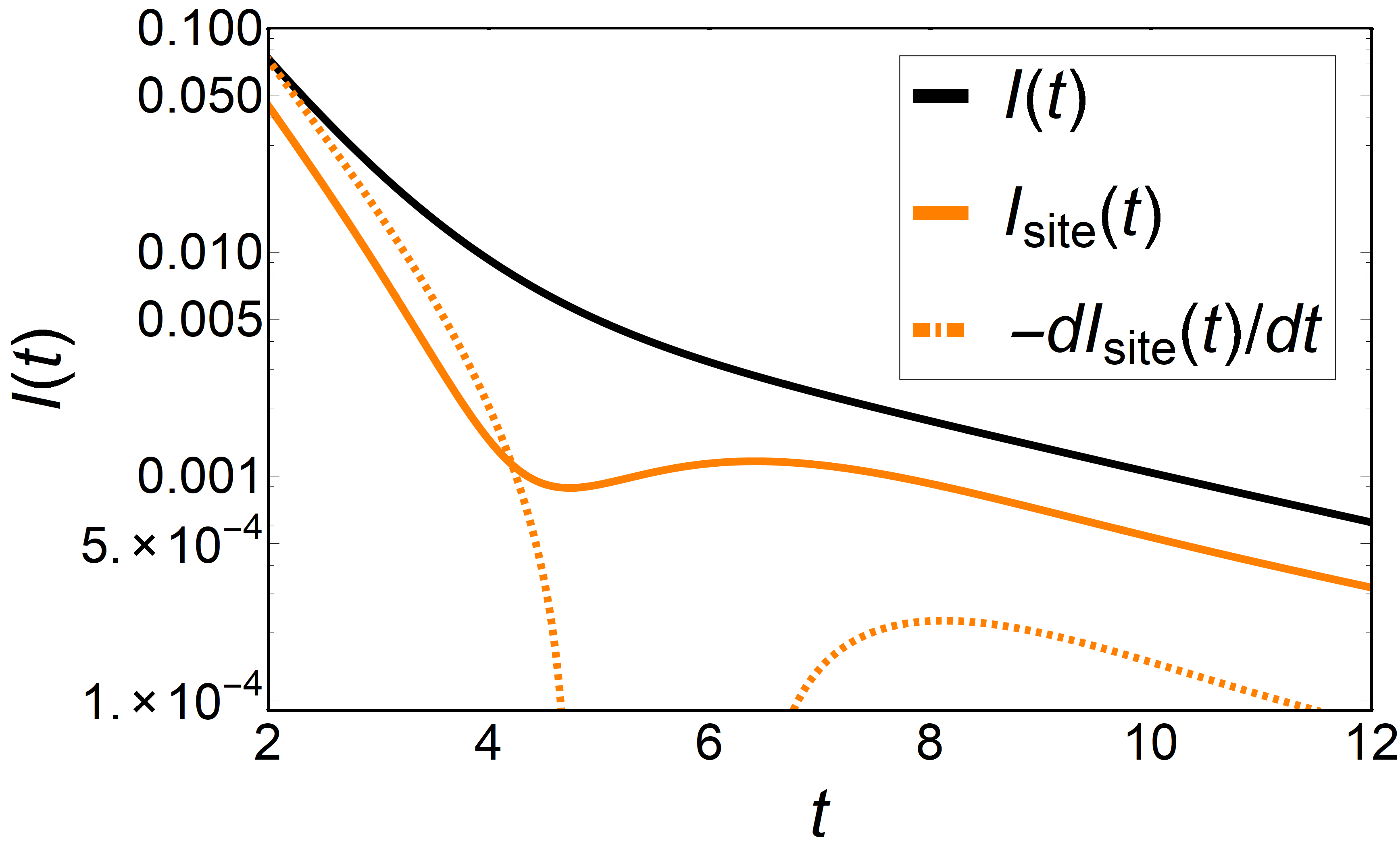}
\caption{A sketch of the Markov-jump model Eq.~\eqref{chain-master} (left) and the corresponding behavior of the Fisher information (right) for $r^* = 0.1$.
As discusussed in Section \ref{sec-monotonicity} and Appendix \ref{app-fisher-markov}, the Fisher information of the detailed probabilities $P_i^\alpha$ and $P_i^\beta$ taking into account both the sites and the internal state is a monotonically decreasing function of time (black).
By contrast, projecting the dynamics onto only the sites yields a non-monotonic behavior of the corresponding Fisher information (orange). For clarity, we also show the negative derivative of this quantity (orange, dashed). \label{fig-hidden-model}}
\end{figure*}
To demonstrate how the Fisher information can be used as a tool to reveal hidden states, we construct a simple Markov jump model.
We imagine a small machine (e.~g.~a molecular motor), which can move between $N$ sites on a one-dimensional chain, which are indexed by $i$.
For simplicity, we only allow moves to neighboring sites, i.~e.~from $i$ to $i+1$ or $i-1$.
We further assume that the machine has two internal states, one inactive state $\alpha$, in which the transition rates to the left and right are small and symmetric (e.~g.~driven by thermal noise), $W^\alpha_{i+1,i} = W^\alpha_{i-1,i} = r^\alpha/2$.
The other state $\beta$ is an active state, where the rate of right and left moves are different $W^\beta_{i+1,i} = r^\beta q, W^\beta_{i-1,i} = r^\beta(1-q)$, where the parameter $0 < q < 1$ determines the asymmetry between the rates.
Finally, the rate of change of the internal state is $r^* s$ from $\alpha$ to $\beta$ and $r^* (1-s)$ from $\beta$ to $\alpha$, with $0 < s < 1$.
This system is sketched in Fig.~\ref{fig-hidden-model}.
We note that similar, albeit more detailed models have been used to study the energetics of actual molecular motors \cite{Gas07}; in this case, $i$ corresponds to the position of the motor, whereas the internal state represents different chemical states.
Denoting the probability to be at site $i$ in state $\alpha$ at time $t$ as $P_i^\alpha(t)$, the evolution of these probabilities is given by the master equation
\begin{subequations}
\begin{align}
    d_t P_i^\alpha &= r^\alpha \big( \frac{1}{2} P_{i-1}^\alpha + \frac{1}{2} P_{i+1}^\alpha - P_{i}^\alpha \big) \\
    & \qquad + r^*\big((1-s) P_i^\beta - s P_i^\alpha \big) \nn
    d_t P_i^\beta &= r^\beta \big( q P_{i-1}^\beta + (1-q) P_{i+1}^\beta - P_i^\beta \big) \\
    & \qquad + r^* \big(s P_i^\alpha - (1-s) P_i^\beta \big) \n .
\end{align}\label{chain-master}%
\end{subequations}
We consider the situation where the position of the machine is the observable, while its internal state cannot be directly observed and thus constitutes a hidden degree of freedom.
The total (observable) probability to be at site $i$ is $P_i = P_i^\alpha + P_i^\beta$.
The question we wish to answer is whether, by observing only the position and thus $P_i$, we can draw any conclusions about the presence of the hidden internal states of the machine.
Under periodic boundary conditions, i.~e.~identifying the sites $N+1$ and $1$, respectively $N$ and $0$ with each other, a straightforward calculation shows that in the steady state, the probabilities are given by $\pi_i = 1/N$ and we have a current $J = (2q-1)s r_\beta/N$ flowing between any two sites $i$ and $i+1$.
However, neither of these steady state quantities necessarily indicates the presence of the internal states, since the effective master equation
\begin{align}
    d_t P_i = s r^\beta \big( q P_{i-1} + (1-q) P_{i+1} - P_{i} \big) \label{master-chain-eff}
\end{align}
with no internal states leads to the same probabilities and current in the steady state.
However, the time evolution resulting from Eqs.~\eqref{chain-master} and \eqref{master-chain-eff} starting from an arbitrary initial state is generally different.
In particular, the Fisher information $I_\text{site} = \sum_{i=1}^N (d_t \ln P_i)^2 P_i$ obtained from observing only the occupation probabilities of the sites is a monotonically decreasing function for the effective Eq.~\eqref{master-chain-eff}, see Appendix \ref{app-fisher-markov}.
By contrast, for Eq.~\eqref{chain-master}, the evolution of $P_i$ generally depends on the internal state and the Fisher information may exhibit a non-monotonic behavior.
The time derivative of $I_\text{site}$ is given by
\begin{widetext}
\begin{align}
    d_t I_\text{site} = \sum_{i=1}^N \bigg[ &\Big( (r^\alpha-r^\beta) \big( \dot{\kappa}_i^+ + \dot{\kappa}_i^- \big) - r^\beta (2q-1) \big(\dot{\kappa}_i^+ - \dot{\kappa}_i^-\big) + 2 \big(r^*(2s-1) + r^\alpha-r^\beta \big) \dot{\Phi}_i \Big) \dot{\nu}_i \label{fisher-chain-site}\\
    &\qquad - \frac{r^\alpha}{2} \nu_i \big( (\dot{\kappa}_i^+)^2 + (\dot{\kappa}_i^-)^2 \big) - r^\beta (1-\nu_i) \big( q (\dot{\kappa}_i^+)^2 + (1-q) (\dot{\kappa}_i^-)^2 \big) \bigg] P_i \n ,
\end{align}
\end{widetext}
where we defined $\Phi_i = \ln (P_i)$, $\kappa_i^{\pm} = \ln(P_{i \pm 1}/P_i)$ and $\nu_i = P_i^\alpha/P_i$ and $\dot{\Phi}$ denotes a time derivative.
Since $0 < \nu_i < 1$, the two terms in the second line are negative; they describe the relaxation of the site occupation probability towards a uniform probability.
By contrast, the terms in the first line can be positive or negative.
Importantly, they are non-zero only if the ratio $\nu_i = P_i^\alpha/P_i$ changes in time, i.~e.~if the relative probability to be in state $\alpha$ at site $i$ changes.
This is equivalent to the condition formulated in Section \ref{sec-monotonicity} that the conditional probability of the hidden degrees of freedom should change with time.
As a concrete example, we take a periodic chain consisting of $N=5$ sites and set $s=1/2$, $r^\alpha=0.1$, $r^\beta=1$ and $q=0.9$.
This corresponds to a situation where the transitions between the internal states are symmetric, the dynamics are slow in the inactive state and, by contrast, fast and highly directed in the active state.
Since we consider the internal states as hidden, we assume that we do not have any control over their initial preparation.
Thus, as our initial state, we take the machine to be confined to sites $1$ and $2$ and otherwise in the steady state of the dynamics Eq.~\eqref{chain-master}, i.~e.~the steady-state solution to Eq.~\eqref{chain-master} involving only states $1$ and $2$ without periodic boundary conditions.
As shown in Fig.~\ref{fig-hidden-model}, for this initial condition, we indeed find that $I_\text{site}$ is a non-monotonic function of time.
Thus, by observing only the occupation probabilities of the sites and computing the corresponding Fisher information, we can infer the presence of hidden states in the system.
We remark that whether the Fisher information actually exhibits a non-monotonic behavior is highly sensitive to the parameters of the system and its initial preparation.
For example, for the parameters and initial state given above, we only observe an increase of the Fisher information for $r^* \lesssim 0.166$, i.~e.~if the transitions between the internal states are sufficiently slow.
Further, for the above parameters, we never observe an increase in the Fisher information if we initialize the system at a single site, as the negative contribution from the relaxation of the site occupation in Eq.~\eqref{fisher-chain-site} always dominates.
However, this is not a generic feature, for example for $r^\alpha = r^\beta = 1$ and $r^* = 0.1$, an increasing Fisher information can be observed even starting from a single site.

\section{Discussion} \label{sec-discussion}

The speed limit Eq.~\eqref{speed-limit-0} on the time-evolution of the average of a fluctuating observable shows that the behavior of measurable observables (averages and fluctuations) is governed by the information-theoretic concept of Fisher information.
A similar connection between the Fisher information and the family of thermodynamic uncertainty relations was recently obtained in Refs.~\cite{Has18,Dec18C}.
Such a connection can potentially be exploited in several ways. 
If the underlying probability distribution and the corresponding Fisher information is not known, then we can obtain a lower bound in terms of measurable quantities.
Since the lower bound is guaranteed to hold for all observables, we may also compare the bounds obtained by measuring different observables in order to find the observable that contains the most information about the time-evolution of the probability density.

On the other hand, if we have a theoretical model for a particular physical system, then the speed limit can serve as a test for the validity of the model: If we find that the observed time evolution of any observable exceeds the Fisher information bound predicted by the theoretical model, then this is a sure indication that crucial information about the system is missing in the model.
For systems without explicit time-dependence the monotonic decay of the Fisher information provides even stricter restrictions on the type of models that can describe a given system.
Finally, if the Fisher information itself is known, then the speed limit imposes a regularity condition on the system in the sense that it limits the rate of change of any conceivable observable.

We remark that trade-off-relations between speed, accuracy and the cost of driving the system have been observed in many contexts, both theoretically and experimentally \cite{Lan12,Lah16,Bar16,Pie18}.
The speed limits Eq.~\eqref{speed-limit-0} shows that such trade-off-relations exist independently of the precise nature of the stochastic dynamics and are a consequence of information-theoretic bounds.
This reinforces the insight that the information content of a physical system has measurable consequences and that information theory can be a useful tool to characterize the properties of the system.

\begin{acknowledgments}
\textbf{Acknowledgments.} This work was supported by the World Premier International Research Center Initiative (WPI), MEXT, Japan. 
S.~I.~is supported by JSPS KAKENHI Grant No. JP16K17780 and JP19H05796, and JST Presto Grant Number JP18070368, Japan.
A.~D.~is supported by JSPS Grant-in-Aid for Scientific Research on Innovative Areas "Discrete Geometric Analysis for Materials Design": Grant Number 17H06460. 
The authors wish to thank S.-i.~Sasa for helpful discussions.
\end{acknowledgments}

\bibliography{bib}

\onecolumngrid
\appendix

\section{Sufficient statistic for time evolution} \label{app-sufficient}
In Section \ref{sec-intrinsic-speed}, we posed the question of what is the probability density that gives rise to equality in Eq.~\eqref{speed-limit-0}.
To answer this, we briefly review the derivation of the Cram{\'e}r-Rao bound for the pertinent case.
For some observable $r(\bm{x})$, we can write the time derivative of its average as
\begin{align}
d_t \av{r}_t = \int d\bm{x} \ r(\bm{x}) \partial_t P(\bm{x},t) = \int d\bm{x} \ \big(r(\bm{x}) - \av{r}_t \big) \partial_t \ln \big(P(\bm{x},t)\big) P(\bm{x},t) ,
\end{align}
where we used that the average of $\partial_t \ln (P(\bm{x},t))$ is zero due to conservation of probability.
Taking the square and applying the Cauchy-Schwarz inequality, we obtain
\begin{align}
\big( d_t \av{r}_t \big)^2 \leq \Av{\big(r-\av{r}_t \big)^2}_t I(t) \label{cramer-rao-2} ,
\end{align}
this is the Cram{\'e}r-Rao bound, which is the square of Eq.~\eqref{speed-limit-0}.
Now in order to have equality in the Cauchy-Schwarz inequality 
\begin{align}
\av{f g}_t^2 \leq \av{f}_t^2 \av{g}_t^2,
\end{align}
either one of the functions should be constant (we ignore this trivial case), or the two functions have to be linearly dependent
\begin{align}
f(x) = \alpha g(x) .
\end{align}
Thus, in order to obtain equality in Eq.~\eqref{cramer-rao-2}, we must have
\begin{align}
\partial_t \ln \big(P(\bm{x},t)\big) = \alpha(t) \big(r(\bm{x}) - \av{r}_t \big) .
\end{align}
This constitutes a differential equation for $P(\bm{x},t)$ with respect to time, whose solution is
\begin{align}
P(\bm{x},t) = e^{\int_0^t ds \ \alpha(s) (r(\bm{x}) - \av{r}_s)} P_0(\bm{x}) .
\end{align}
Defining $A(t) = \int_0^t ds \ \alpha(s)$ and using the fact that $P(\bm{x},t)$ has to be normalized, we can also write this as
\begin{align}
P(\bm{x},t) = \frac{e^{A(t) r(\bm{x})}}{\av{e^{A(t) r(\bm{x})}}_0} P_0(\bm{x}) 
\end{align}
with boundary condition $A(0)=0$, where $\av{\ldots}_0$ denotes an average with respect to the initial probability density $P_0(\bm{x})$.
All the information about the time evolution of the probability density is encoded in the constant $A(t)$.
Since we have
\begin{align}
\frac{d_t \av{r}_t}{\av{\Delta r^2}_t} = d_t A(t) ,
\end{align}
we can infer the time evolution of the probability density by measuring the average and variance of the observable $r$.
In that sense, while $r(\bm{x})$ is not a sufficient statistic of $P(\bm{x},t)$ (since we cannot generally infer $P_0(\bm{x})$ from knowledge about $r$), the time evolution of $r$ contains all the information about the time evolution of $P(\bm{x},t)$.
Thus, $r$ may be viewed as a sufficient statistic for the time evolution of the probability density.
Explicitly, we can write
\begin{align}
P(\bm{x},t) &= \frac{\exp\Big[{\int_0^t ds \ \frac{d_{s} \av{r}_{s}}{\av{\Delta r^2}_s}  r(\bm{x})} \Big]}{\Av{\exp\Big[{\int_0^t ds \ \frac{d_s \av{r}_s}{\av{\Delta r^2}_s}  r(\bm{x})} \Big]}_0} P_0(\bm{x}) \nn
&= \frac{\exp\Big[\Big( \frac{\av{r}_t}{\av{\Delta r^2}_t} - \frac{\av{r}_0}{\av{\Delta r^2}_0} + \int_0^t ds \ \frac{\av{r}_{s}}{\av{\Delta r^2}_s} d_s \ln \big(\av{\Delta r^2}_s\big)\Big)  r(\bm{x}) \Big]}{\Av{\exp\Big[\Big( \frac{\av{r}_t}{\av{\Delta r^2}_t} - \frac{\av{r}_0}{\av{\Delta r^2}_0} + \int_0^t ds \ \frac{\av{r}_{s}}{\av{\Delta r^2}_s} d_s \ln \big(\av{\Delta r^2}_s\big)\Big)  r(\bm{x}) \Big]}_0} P_0(\bm{x}),
\end{align}
which makes the dependence of the probability density on the average and variance of $r$ explicit.
If the variance of $r$ is independent of time then this simplifies to
\begin{align}
P(\bm{x},t) &= \frac{\exp\Big[ \frac{(\av{r}_t - \av{r}_0)r(\bm{x})}{\av{\Delta r^2}}\Big]}{\Av{\exp\Big[ \frac{(\av{r}_t - \av{r}_0)r(\bm{x})}{\av{\Delta r^2}}\Big]}_0} P_0(\bm{x}) \label{equality-cond} .
\end{align}
As an example of a Fokker-Planck dynamics which explicitly realizes the above probability distribution, we consider an overdamped particle in a parabolic trap with spring constant $\kappa$, centered at the time-dependent position $a(t)$,
\begin{align}
\partial_t P(x,t) = \mu \partial_x \big( \kappa (x - a(t)) + \kb T \partial_x \big) P(x,t).
\end{align}
If the variance of the particle's position initially has the equilibrium value $\av{\Delta x^2}_\text{eq} = T/\kappa$, then it will remain constant in time.
Since the the force is linear in $x$, the distribution is further Gaussian and can be written as
\begin{align}
P(x,t) &= \frac{1}{\sqrt{2 \pi \av{\Delta x^2}_\text{eq}}} e^{-\frac{(x-\av{x}_t)^2}{2\av{\Delta x^2}_\text{eq}}} = e^{\frac{(\av{x}_t - \av{x}_0) x}{\av{\Delta x^2}_\text{eq}}} e^{-\frac{(\av{x}_t)^2 - (\av{x}_0)^2}{\av{\Delta x^2}_\text{eq}}} \frac{1}{\sqrt{2 \pi \av{\Delta x^2}_\text{eq}}} e^{-\frac{(x-\av{x}_0)^2}{2\av{\Delta x^2}_\text{eq}}} \nn
&=  \frac{\exp\Big[\frac{(\av{x}_t - \av{x}_0) x}{\av{\Delta x^2}_\text{eq}}\Big]}{\Av{\exp\Big[\frac{(\av{x}_t - \av{x}_0) x}{\av{\Delta x^2}_\text{eq}}\Big]}_0} P_0(x) ,
\end{align}
which is precisely of the form Eq.~\eqref{equality-cond}  with $r(x) = x$.
We thus have equality in Eq.~\eqref{speed-limit-0},
\begin{align}
\frac{\vert d_t \av{x}_t \vert}{\sqrt{\av{\Delta x^2}_\text{eq}}} = v_I(t) .
\end{align}

\section{Thermodynamic cost for Fokker-Planck dynamics} \label{app-cost-fp}

In Ref.~\cite{Ito17}, the justification for referring to the quantity $\mathcal{C}$ as a thermodynamic cost was provided by relating it to the entropy change upon the system transitioning between two discrete states $x$ and $x'$.
To provide the analog for the case when the system is described by a set of continuous variables, we first note that the Fokker-Planck equation \eqref{fokkerplanck} for the probability density is equivalent to the stochastic evolution of the state $\bm{x}(t)$ of the system described by the Langevin equation \cite{Ris86}
\begin{align}
d\bm{x}(t) = \bm{a}(\bm{x}(t),t)dt + \sqrt{2 \bm{B}(\bm{x}(t),t)} \cdot d\bm{W}(t) \label{langevin},
\end{align}
where $\sqrt{\bm{B}}$ refers to the unique positive semidefinite principal square root of the symmetric and positive semidefinite matrix $\bm{B}$.
$\bm{W}$ is a vector of mutually uncorrelated Wiener processes and $\cdot$ denotes the It{\=o} product.
We remark that the form Eq.~\eqref{langevin} is more general than Eq.~\eqref{langevin-0}, which is obtained by setting $\bm{a} = \bm{\mu} \bm{f}$ and $\bm{B} = \bm{\mu} T$.
We want to describe the stochastic Shannon entropy (or generalized potential)
\begin{align}
\Phi^\text{sys}(t) = - \ln P(\bm{x}(t),t),
\end{align}
with $\Av{\Phi^\text{sys}}_t = S^\text{sys}(t)$.
We rewrite Fokker-Planck equation \eqref{fokkerplanck} as a continuity equation in terms of the probability current $\bm{j}(\bm{x},t)$,
\begin{align}
\partial_t P(\bm{x},t) &= -\bm{\nabla} \bm{j}(\bm{x},t) \quad \text{with} \label{fokkerplanck-2} \quad
\bm{j}(\bm{x},t) = \Big( \bm{a}(\bm{x},t) - \bm{\nabla} \bm{B}(\bm{x},t) \Big) P(\bm{x},t) ,
\end{align}
where we define the operator $\bm{\nabla} \bm{B} = \partial_{x_j} B_{i j}$.
By It{\=o}'s Lemma, we have for the differential of $\Phi^\text{sys}$,
\begin{align}
d \Phi^\text{sys}(t) &= -\partial_t \ln P(\bm{x}(t),t) dt - \bm{\nabla} \ln P(\bm{x}(t),t) \cdot d\bm{x}(t) 
 - \bm{B}(\bm{x},t) \bm{\nabla} \bm{\nabla} \ln P(\bm{x},t)  ,
\end{align}
where by $\bm{B} \bm{\nabla} \bm{\nabla}$ we mean the operator $B_{i j} \partial_{x_i} \partial_{x j}$.
We can equivalently write this using the Stratonovich product $\circ$,
\begin{align}
d \Phi^\text{sys}(t) = -\partial_t \ln P(\bm{x}(t),t) dt - \bm{\nabla} \ln P(\bm{x}(t),t) \circ d\bm{x}(t).
\end{align}
The first term describes the change in Shannon entropy in a fixed state $\bm{x}$ due to the change in the ensemble probability $P(\bm{x},t)$ to be in state $\bm{x}$.
We interpret this term as a \textit{global} (in the sense of ensemble) contribution; note that due to conservation of probability, this term always vanishes on average.
On the other hand, the second, \textit{local}, contribution describes the change in Shannon entropy due to a change in state from $\bm{x}$ to $\bm{x}' = \bm{x} + d\bm{x}$; this change in Shannon entropy in a transition $\Delta \sigma^\text{sys}_{x' \rightarrow x}$ of a Markov jump process, as defined in Ref.~\cite{Ito17}.
In analogy to Ref.~\cite{Ito17}, we thus interpret 
\begin{align}
\Delta\bm{\Sigma}^\text{sys}_\text{loc}(\bm{x},t) \equiv - \bm{\nabla} \ln P(\bm{x},t)  \label{shannon-loc}
\end{align}
as the \textit{local} change in Shannon entropy, which is related to the change in average Shannon entropy via
\begin{align}
d \Sigma^\text{sys} = \Av{ \Delta\bm{\Sigma}^\text{sys}_\text{loc} \circ d\bm{x}} = \int d\bm{x} \ \Delta\bm{\Sigma}^\text{sys}_\text{loc}(\bm{x},t) \bm{j}(\bm{x},t) dt 
\end{align} 
Using this definition and integrating by parts, it is then easy to show that
\begin{align}
-\big\langle\partial_t \Delta\bm{\Sigma}^\text{sys}_\text{loc} \circ \dot{\bm{x}}\big\rangle_t &\equiv -\int d\bm{x} \  \bm{j}(\bm{x},t) \partial_t \Delta\bm{\Sigma}^\text{sys}_\text{loc}(\bm{x},t)  = \int d\bm{x} \ \bm{j}(\bm{x},t) \bm{\nabla} \partial_t \ln P(\bm{x},t) = - \int d\bm{x} \ \partial_t \ln P(\bm{x},t) \bm{\nabla}  \bm{j}(\bm{x},t) \nn
&  = \int d\bm{x} \ \partial_t \ln P(\bm{x},t) \partial_t P(\bm{x},t) = \bigg(\frac{ds}{dt}\bigg)^2  \label{fisher-langevin} ,
\end{align}
in analogy to Eq.~(37) of Ref.~\cite{Ito17}.
For a diagonal diffusion matrix $B_{i j} = B_i \delta_{i j}$ with $B_i > 0$, we can further write the change in total entropy as follows \cite{Che06,Spi12},
\begin{align}
d\Phi^\text{tot}(t) &= d\Phi^\text{sys}(t) + d\Phi^\text{med}(t) \quad \text{with} \quad
d\Phi^\text{med}(t) = \big(B_i(\bm{x}(t),t)\big)^{-1} \Big( a_i(\bm{x}(t),t)  - \partial_{x_i} B_i(\bm{x}(t),t) \Big) \circ dx_i(t)  .
\end{align}
Defining the local change in medium entropy and total entropy
\begin{align}
\Delta\bm{\Sigma}^\text{med}_\text{loc}(\bm{x},t) &\equiv \Big(\bm{a}(\bm{x},t) - \bm{b}'(\bm{x},t) \Big) \bm{B}^{-1}(\bm{x},t) \\
\Delta\bm{\Sigma}^\text{tot}_\text{loc}(\bm{x},t) &\equiv \Delta\bm{\Sigma}^\text{med}_\text{loc}(\bm{x},t) + \Delta\bm{\Sigma}^\text{sys}_\text{loc}(\bm{x},t) \n
\end{align}
with the vector $b_i'(\bm{x},t) = \partial_{x_i} B_{i}(\bm{x},t)$, we thus have for the average change in total entropy \cite{Spi12}
\begin{align}
d\Sigma^\text{tot} &= \Av{\Big(\Delta\bm{\Sigma}^\text{med}_\text{loc} + \Delta\bm{\Sigma}^\text{sys}_\text{loc} \Big) \circ d\bm{x}} = \int d\bm{x} \ \frac{\bm{j}(\bm{x},t) \bm{B}^{-1}(\bm{x},t) \bm{j}(\bm{x},t)}{P(\bm{x},t)} dt  .
\end{align}
This further allows us to write
\begin{align}
\bigg(\frac{ds}{dt}\bigg)^2 = \Av{\Big(\partial_t \Delta\bm{\Sigma}^\text{med}_\text{loc} - \partial_t \Delta\bm{\Sigma}^\text{tot}_\text{loc}\Big) \circ \dot{\bm{x}}},
\end{align}
again in analogy to the identification made in Ref.~\cite{Ito17}.

\section{Monotonicity of Fisher information} \label{app-fisher-markov}

Here, we prove the decomposition of the Fisher information given in Eq.~\eqref{fisher-split}, for the specific cases of a Fokker-Planck and Markov jump dynamics.
We now assume that $P(\bm{x},t)$ describes the time-evolution of a diffusive dynamics, i.~e.~is the solution of the Fokker-Planck equation corresponding to Eq.~\eqref{langevin} \cite{Ris86}
\begin{align}
\partial_t P(\bm{x},t) &= \mathcal{G}(\bm{x},t) P(\bm{x},t) \qquad \text{with}  \qquad \mathcal{G}(\bm{x},t) = - \partial_{x_i} \Big( a_i(\bm{x},t) - \partial_{x_j} B_{i j}(\bm{x},t) \Big) \label{fokkerplanck-general}  ,
\end{align}
where a sum over repeated indices is implied.
Here $\bm{a}(\bm{x},t)$ is a drift vector and $\bm{B}(\bm{x},t)$ is a symmetric and positive semidefinite diffusion matrix, i.~e.
\begin{align}
v_i B_{i j}(\bm{x},t) v_j \geq 0
\end{align}
for an arbitrary vector $\bm{v}$ and for all $\bm{x}$ and $t$.
Note that this form includes Eq.~\eqref{fokkerplanck} as a special case for $a_i = \mu_{i j} f_j$ and $B_{i j} = \kb T  \mu_{i j}$.
The Fokker-Planck operator $\mathcal{G}$ is the generator of the dynamics.
We further introduce the adjoint of the generator,
\begin{align}
\mathcal{G}^\dagger(\bm{x},t) = \Big(a_i(\bm{x},t) + B_{i j}(\bm{x},t) \partial_{x_j} \Big)  \partial_{x_i},
\end{align}
which satisfies
\begin{subequations}
\begin{align}
\int d\bm{x} \ f \mathcal{G} g &= \int d\bm{x} \ g \mathcal{G}^\dagger f \\
\mathcal{G}^\dagger f^2 &= 2 f \mathcal{G}^\dagger f + 2 \big[\partial_{x_i} f \big] B_{i j} \big[\partial_{x_j} f \big]
\end{align}
\end{subequations}
for suitable (smooth and integrable) functions $f(\bm{x},t)$ and $g(\bm{x},t)$.
For such a dynamics, we consider the time-derivative of the Fisher information
\begin{align}
d_t I(t) = \int d\bm{x} \ \frac{2 P \big[\partial_t P \big] \big[\partial_t^2 P \big] - \big[ \partial_t P \big]^3}{P^2} ,
\end{align}
with the convention that derivatives inside square brackets do not act on terms outside the brackets.
Here and in the following, we omit the arguments of the respective functions for brevity.
We write the second time-derivative of the probability density as $\partial_t^2 P = \partial_t \mathcal{G} P = \dot{\mathcal{G}} P + \mathcal{G} \partial_t P$, where we introduced the time-derivative of the Fokker-Planck operator
\begin{align}
\dot{\mathcal{G}}(\bm{x},t) = - \partial_{x_i} \Big( \big[ \partial_t a_i(\bm{x},t) \big] - \partial_{x_j} \big[ \partial_t B_{i j}(\bm{x},t) \big] \Big) .
\end{align}
Defining the generalized potential $\Phi(\bm{x},t) = -\ln (P(\bm{x},t))$, which can be identified as a stochastic Shannon entropy in the sense that the Shannon entropy is the average of $\Phi$, $\Sigma^\text{sys} = -\int d\bm{x} \ \ln(P) P = \av{\Phi}_t$, we can write for the time-derivative of the Fisher information
\begin{align}
d_t I(t) &+ 2 \int d\bm{x} \ \big[\partial_t \Phi \big] \dot{\mathcal{G}} P = - \int d\bm{x} \ \bigg( 2  \big[ \partial_t \phi \big] \mathcal{G}^2 P + \big[ \partial_t \Phi \big]^2 \mathcal{G} P \bigg) \nn
&= - \int d\bm{x} \ \bigg( 2  \big[\mathcal{G} P \big] \big[ \mathcal{G}^\dagger \partial_t \Phi \big] + P \big[ \mathcal{G}^\dagger (\partial_t \Phi)^2 \big] \bigg) \nn
&= -2 \int d\bm{x} \ \bigg( \big[\mathcal{G} P \big] \big[ \mathcal{G}^\dagger \partial_t \Phi \big] + P \big[ \partial_t \Phi \big] \big[ \mathcal{G}^\dagger \partial_t \Phi \big] + P \big[\partial_{x_i} \partial_t \Phi \big] B_{i j} \big[\partial_{x_j} \partial_t \Phi \big]\bigg) \nn
&= - 2 \int d\bm{x} \ \big[\partial_{x_i} \partial_t \Phi \big] B_{i j} \big[\partial_{x_j} \partial_t \Phi \big] P  - \int d\bm{x} \ \bigg( \big[\mathcal{G} P \big] + P \big[\partial_t \Phi \big] \bigg) \big[ \mathcal{G}^\dagger \partial_t \Phi \big] \n .
\end{align}
From the definition of $\Phi$ we have $P \partial_t \Phi = - \partial_t P = - \mathcal{G} P$, such that the last term vanishes.
We thus arrive at
\begin{align}
d_t I(t) &= \underbrace{-2 \Av{ \big[ \dot{\mathcal{G}}^\dagger \partial_t \Phi \big]}_t}_{d_t I_\text{drv}(t)} \underbrace{-2 \Av{\big[\partial_{x_i} \partial_t \Phi \big] B_{i j} \big[\partial_{x_j} \partial_t \Phi \big]}_t}_{d_t I_\text{rel}(t)} \label{fisher-derivative}
\end{align}
with the operator
\begin{align}
\dot{\mathcal{G}}^\dagger(\bm{x},t) =  \Big( \big[ \partial_t a_i(\bm{x},t) \big] + \big[ \partial_t B_{i j}(\bm{x},t) \big] \partial_{x_j}  \Big) \partial_{x_i} .
\end{align}
This is precisely the decomposition Eq.~\eqref{fisher-split}.

Next, consider a Markov jump process on a set of $M$ discrete states defined by the (generally time-dependent) transition rates $W_{i j}(t) \geq 0$ from state $j$ to state $i$ and occupation probabilities $p_i(t)$ of state $i$.
The time-evolution of the occupation probabilities is governed by the Master equation \cite{Kam92}
\begin{align}
d_t p_i(t) = \sum_j \Big( W_{i j}(t) p_j(t) - W_{j i}(t) p_i(t) \Big) = \sum_j \mathcal{G}_{i j}(t) p_j(t) \label{master},
\end{align}
where we defined the matrix-valued generator $\bm{\mathcal{G}}(t)$
\begin{align}
\mathcal{G}_{i j}(t) = W_{i j}(t) - \delta_{i j} \sum_{k} W_{k i}(t) \label{markov-generator} .
\end{align}
In analogy to the continuous case, the (temporal) Fisher information is given in terms of the time-derivative of the occupation probability \cite{Ito17},
\begin{align}
I(t) = \sum_i \frac{\big(d_t p_i(t)\big)^2}{p_i(t)} .
\end{align}
The time-derivative of the Fisher information is then
\begin{align}
d_t I(t) = \sum_i \frac{2 p_i(t) \big[d_t p_i(t)\big] \big[d_t^2 p_i(t)\big] - \big[d_t p_i(t) \big]^3}{p_i(t)^2} = \sum_i \Big( 2\big[d_t \ln p_i(t)\big] \big[d_t^2 p_i(t) \big] - \big[ d_t \ln p_i(t) \big]^2 \big[ d_t p_i(t) \big] \Big),
\end{align}
or in terms of the generator
\begin{align}
d_t I(t) &= \sum_i \Big( 2\big[d_t \ln p_i(t)\big] \Big[d_t \sum_j \mathcal{G}_{i j}(t) p_j(t) \Big] - \big[ d_t \ln p_i(t) \big]^2 \Big[ \sum_j \mathcal{G}_{i j}(t) p_j(t) \Big] \Big) \\
&= 2 \sum_{i,j} \big[d_t \ln p_i(t)\big] \dot{\mathcal{G}}_{i j}(t) p_j(t) + \sum_{i,j} \Big( 2\big[d_t \ln p_i(t)\big] \mathcal{G}_{i j}(t) \big[ d_t \ln p_j(t)\big] - \big[ d_t \ln p_i(t) \big]^2 \mathcal{G}_{i j}(t) \Big) p_j(t) , \n
\end{align}
where we introduced the time-derivative of the generator $\dot{\bm{\mathcal{G}}}(t)$.
We define $a_i \equiv d_t \ln p_i(t)$, in terms of which we can rewrite the above as
\begin{align}
d_t I(t) = 2 \bm{a}^T \dot{\bm{\mathcal{G}}} \bm{b} + \sum_{i, j} \Big(2 a_i \mathcal{G}_{i j} a_j p_j - a_i^2 \mathcal{G}_{i j} p_j \Big) .
\end{align}
We now plug the explicit definition \eqref{markov-generator} of the generator into the second term,
\begin{align}
\sum_{i, j} \Big(2 a_i \mathcal{G}_{i j} a_j p_j - a_i^2 \mathcal{G}_{i j} p_j \Big) &= \sum_{i,j} \Big( a_i \big( 2 \mathcal{G}_{i j} a_j - a_i \mathcal{G}_{i j} \big) p_j \Big) \\
&= \sum_{i,j} \Big( a_i \big( 2 \big( W_{i j}(t) - \delta_{i j} \sum_{k} W_{k i}(t) \big) a_j - a_i \big( W_{i j}(t) - \delta_{i j} \sum_{k} W_{k i}(t) \big) \big) p_j \Big) \nn
&= \sum_{i,j} \Big( 2 a_i W_{i j} a_j p_j - a_i^2 W_{i j} p_j \Big) - \sum_{i,k} \Big( 2 a_i^2 p_i W_{k i} - a_i^2 p_i W_{k i} \Big) \nn
&= \sum_{i,j} \Big(2 a_i W_{i j} a_j p_j - a_i^2 W_{i j} p_j - W_{i j} a_j^2 p_j  \Big) \nn
&= -\sum_{i,j} \Big( \big(a_i - a_j \big)^2 W_{i j} p_j \Big) \n ,
\end{align}
where we renamed the summation indices in the last term from $(i,k)$ to $(j,i)$ in the second-to-last step.
Since the both the transition rates and occupation probabilities are positive, $W_{i j} \geq 0$ and $p_i \geq 0$, this term is evidently negative.
We thus arrive at
\begin{align}
d_t I(t) = \underbrace{- 2 \big[d_t \bm{\Phi}(t) \big]^T \dot{\bm{\mathcal{G}}}(t) \bm{p}(t) \vphantom{\sum_{i,j} \Big(d_t \Phi_i(t) - d_t \Phi_j(t) \Big)^2}}_{d_t I_\text{drv}(t) } \underbrace{- \sum_{i,j} \Big(d_t \Phi_i(t) - d_t \Phi_j(t) \Big)^2 W_{i j}(t) p_j(t)}_{d_t I_\text{rel}(t)} \label{fisher-derivative-markov} ,
\end{align}
where, in analogy to the continuous case, we introduced the vector of state-dependent Shannon entropy $\bm{\Phi}$ defined by $\Phi_i = - \ln p_i$.
As in Eq.~\eqref{fisher-derivative}, the time-derivative of the Fisher information decomposes into a driving term involving the explicit time-dependence of the generator and a negative semidefinite term, which describes relaxation towards the instantaneous steady state.
If the transition rates do not depend explicitly on time, $d_t W_{i j} = 0$, then, just as in the case of Fokker-Planck dynamics, the Fisher information decreases monotonically in time
\begin{align}
d_t I(t) = d_t I_\text{rel}(t) \leq 0 \label{fisher-monotonic-markov},
\end{align}
in complete analogy to Eq.~\eqref{fisher-monotonic}.
We remark that the same result holds for a mixed process,
\begin{align}
\partial_t P^k(\bm{x},t) = - \partial_{x_i} \Big( a_i^k(\bm{x}) - \partial_{x_j} B_{i j}^k(\bm{x}) \Big) P^k(\bm{x},t) + \sum_{l} \Big( W^{k l}(\bm{x}) P^l(\bm{x},t) - W^{l k}(\bm{x}) P^k(\bm{x},t) \Big) ,
\end{align}
 i.~e.~a Fokker-Planck dynamics with additional discrete states labeled by $k$ and a state-dependent drift vector and diffusion matrix, since the generator is the sum of a diffusion and jump part, to which the arguments leading to Eqs.~\eqref{fisher-monotonic} and \eqref{fisher-monotonic-markov} can be applied separately.

\section{Minimal cost probability density} \label{app-minprob}

Let us consider two particular values $\theta_1$, $\theta_2$ of a parameter and the corresponding probability densities $P^a(\bm{x}) = P(\bm{x},\theta_1)$ and $P^b(\bm{x}) = P(\bm{x},\theta_2)$.
Note that there is an infinite number of possible parameterized probability densities satisfying these conditions, e.~g.~we may have two probability densities $P(\bm{x},\theta)$ and $\tilde{P}(\bm{x},\theta)$ that coincide at $\theta_1$ and $\theta_2$ but are different otherwise.
Each of these possible choices has an associated statistical length and action defined by Eqs.~\eqref{length} and \eqref{action}
\begin{align}
\mathcal{L}(\theta_2, \theta_1) &= \int_{\theta_1}^{\theta_2} d\theta \sqrt{\int d\bm{x} \ \frac{ \big(\partial_\theta P(\bm{x},\theta) \big)^2}{P(\bm{x},\theta)}}, \qquad
\mathcal{C}(\theta_2, \theta_1) = \frac{1}{2} \int_{\theta_1}^{\theta_2} d\theta \int d\bm{x} \ \frac{ \big(\partial_\theta P(\bm{x},\theta) \big)^2}{P(\bm{x},\theta)} ,
\end{align}
where we assumed $\theta_2 > \theta_1$ without loss of generality.
Note that for different $P$ and $\tilde{P}$, also the length and cost are generally different.
However, there exists a unique choice $P^*(\bm{x},\theta)$ which simultaneously minimizes the length and cost.
To see this, we first minimize the cost $\mathcal{C}$ with respect to $P(\bm{x},\theta)$. 
In order to simplify the notation, we first reparameterize $\theta(q) = \theta_2 q + \theta_1 (1-q)$ with $q \in [0,1]$.
Using this, we can write the length and cost as
\begin{align}
\mathcal{L}(\theta_2, \theta_1) &= \int_{0}^{1} dq \sqrt{\int d\bm{x} \ \frac{ \big(\partial_q P(\bm{x},q) \big)^2}{P(\bm{x},q)}} \\
 \mathcal{C}(\theta_2, \theta_1) &= \frac{1}{2 (\theta_2-\theta_1)} \int_{0}^{1} dq \int d\bm{x} \ \frac{ \big(\partial_q P(\bm{x},q) \big)^2}{P(\bm{x},q)} \n  ,
\end{align}
with $P(\bm{x},q) \equiv P(\bm{x},\theta(q))$.
We now want to minimize $\mathcal{C}$ with respect to $P(\bm{x},q)$, under the condition that $P(\bm{x},q)$ is a well-defined probability density, i.~e.~$P(\bm{x},q) > 0$ and $\int d\bm{x} \ P(\bm{x},q) = 1$.
Introducing the Lagrange multiplier $\alpha$, we thus have to minimize the auxiliary functional
\begin{align}
F_{\mathcal{C}}&[P,\partial_q P] = \int_0^1 dq \ f_{\mathcal{C}}[P,\partial_q P](q) \equiv \int_0^1 dq \ \Bigg( \int d\bm{x} \ \frac{ \big(\partial_q P \big)^2}{P} - 4 \alpha \bigg( \int d\bm{x} \ P - 1 \bigg) \Bigg) ,
\end{align}
where the factor $4$ in front of $\alpha$ is included for later notational convenience.
The corresponding Euler-Lagrange equation reads
\begin{align}
\partial_{P} f_{\mathcal{C}} - d_q \partial_{\partial_q P} f_{\mathcal{C}} = \frac{\big(\partial_q P \big)^2}{P^2} - 2 \frac{\partial_q^2 P}{P} - 4 \alpha = 0,
\end{align}
Since $P(\bm{x},q) > 0$, we can write this as
\begin{align}
\big(\partial_q P \big)^2 - 2 P \partial_q^2 P - 4 \alpha P^2 = 0,
\end{align}
which has the general solution
\begin{align}
P(\bm{x},q) = f(\bm{x}) \cos\Big(\sqrt{\alpha} (q - g(\bm{x}) \Big)^2 .
\end{align}
The functions $f(\bm{x})$ and $g(\bm{x})$, as well as the value of $\alpha$ are fixed by the boundary conditions $P(\bm{x},0) = P^a(\bm{x})$ and $P(\bm{x},1) = P^b(\bm{x})$ and the normalization.
The final result for $P^*(\bm{x},q)$ minimizing the cost reads,
\begin{align}
P^*(\bm{x},q) &= \frac{1}{1 - \cos\big(\frac{\Lambda}{2}\big)^2} \Big( \sin\Big(\frac{\Lambda}{2}(1-q)\Big) \sqrt{P^a \vphantom{P^b} (\bm{x})} + \sin\Big(\frac{\Lambda}{2} q\Big) \sqrt{P^b(\bm{x})} \Big)^2 \label{optimal-prob-2} \\
&\text{with} \quad \Lambda = 2 \arccos \bigg( \int d\bm{x} \ \sqrt{P^a(\bm{x}) P^b(\bm{x})} \bigg) \n  .
\end{align}
For this choice, we have $I^*(q) = \int d\bm{x} \ (\partial_q P^*(\bm{x},q))^2/P^*(\bm{x},q) = \Lambda^2$ and thus the minimal cost and statistical length
\begin{align}
\mathcal{C}^* = \frac{\Lambda^2}{2(\theta_2 - \theta_1)}, \qquad \qquad \mathcal{L}^* = \Lambda \label{optimal-cost} .
\end{align}
In hindsight, it is obvious that $\mathcal{C}$ is minimized by a probability density that yields constant Fisher information, since the former is defined as $\mathcal{C} = \int_{\theta_1}^{\theta_2} d\theta \ I(\theta)$.
The same is true for the length $\mathcal{L}$, which is thus also minimized by $P^*$.
We note that, in analogy to the discussion in Ref.~\cite{Ito17}, the choice $P^*(\bm{x},q)$ is the geodesic curve connecting $P^a(\bm{x})$ and $P^b(\bm{x})$, however, the geometric analogy is now less intuitive, since the underlying space is infinite-dimensional.
Since $P^*(\bm{x},q)$ yields the minimal length between $P^a$ and $P^b$ for a normalized probability density, we can interpret $\mathcal{L}^* = \Lambda$ as the arc length between $P^a$ and $P^b$ on the infinite-dimensional unit sphere.

Since $\mathcal{C}^*$ is the minimal cost, any other normalized probability density $\tilde{P}(\bm{x},q)$ results in a larger cost $\tilde{\mathcal{C}} \geq \mathcal{C}^*$.
In particular, for a simple linear interpolation
\begin{align}
\tilde{P}(\bm{x},q) = P^b(\bm{x}) q + P^a(\bm{x}) (1-q),
\end{align}
which is positive and normalized, we obtain the cost
\begin{align}
\mathcal{\tilde{C}}(\theta_2,\theta_1) &= \frac{1}{2 (\theta_2-\theta_1)} \int d\bm{x} \ \big(P^b - P^a \big) \ln \bigg(\frac{P^b}{P^a} \bigg)\\
& = \frac{1}{2(\theta_2 - \theta_1)} \Big( D_\text{KL}(P^b \Vert P^a) + D_\text{KL}(P^a \Vert P^b) \Big) \equiv \frac{1}{\theta_2-\theta_1} D_\text{KL}^\text{sym}(P^b,P^a) \n ,
\end{align}
where we defined the symmetrized Kullback-Leibler divergence or relative entropy.
We thus obtain the by no means obvious lower bound on the latter,
\begin{align}
D_\text{KL}^\text{sym}(P^b,P^a) \geq 2 \arccos \bigg( \int d\bm{x} \ \sqrt{P^a(\bm{x}) P^b(\bm{x})} \bigg)^2 .
\end{align}

Applying the above discussion to the time evolution of a stochastic dynamics $\theta = t$, we fix the initial and final state of the system, $P(\bm{x},0) = P^\text{i}(\bm{x})$ and $P(\bm{x},\mathcal{T}) = P^\text{f}(\bm{x})$.
The optimal time evolution between these two states is given by Eq.~\eqref{optimal-prob-2} with $s = t/\mathcal{T}$.
Since this results in $\mathcal{L}^* = \Lambda$ and $\mathcal{C}^* = \Lambda^2/(2 \mathcal{T})$, we obtain a lower bound on the thermodynamic cost of the evolution from the initial to the final state \cite{Ito17},
\begin{align}
\mathcal{C} \geq \frac{\Lambda^2}{2\mathcal{T}} \quad \text{with} \quad \Lambda = 2\arccos \bigg( \int d\bm{x} \ \sqrt{P^\text{i}(\bm{x}) P^\text{f}(\bm{x})} \bigg) .
\end{align}
Thus, the minimal thermodynamic cost is given by the square of the shortest distance between the initial and final state, divided by the evolution time.
This shows that, in particular, a faster evolution is generally associated with a larger thermodynamic cost; further, zero cost is only realizable in the quasistatic limit where the time evolution is infinitely slow.

\section{Shannon entropy and Fisher information for normal distributions} \label{app-logvar-normal}

We consider a multivariate normal distribution
\begin{align}
P(\bm{x}) = \frac{1}{\sqrt{(2 \pi)^M \det(\bm{\Xi}^{-1})}} \exp \bigg[-\frac{1}{2} \big(\bm{x} - \bm{\mu}\big)^T \bm{\Xi}^{-1} \big(\bm{x} - \bm{\mu}\big) \bigg],
\end{align}
where $M$ denotes the dimension of $\bm{x}$, $\bm{\Xi}$ is the (positive definite and symmetric) covariance matrix defined by
\begin{align}
\Xi_{i j} = \Av{(x_i - \mu_i) (x_j - \mu_j)}
\end{align}
and $\bm{\mu}$ is the average of $\bm{x}$.
We want to compute the variance of the logarithm of $P$,
\begin{align}
\Delta_{\ln} \equiv \Av{(\ln P)^2}-\Av{\ln P \vphantom{)^2}}^2 .
\end{align}
By definition, we have
\begin{align}
\ln (P(\bm{x})) = - \frac{1}{2} \bigg( M \ln (2\pi) + \ln(\det{\bm{\Xi}^{-1}}) + \big(\bm{x} - \bm{\mu}\big)^T \bm{\Xi}^{-1} \big(\bm{x} - \bm{\mu}\big) \bigg) .
\end{align}
Since the first two terms are independent of $\bm{x}$, they do not contribute to the variance and we thus have
\begin{align}
\Delta_{\ln} = \frac{1}{4} \bigg( \Av{\Big(\big(\bm{x} - \bm{\mu}\big)^T \bm{\Xi}^{-1} \big(\bm{x} - \bm{\mu}\big) \Big)^2} - \Av{\big(\bm{x} - \bm{\mu}\big)^T \bm{\Xi}^{-1} \big(\bm{x} - \bm{\mu}\big)}^2 \bigg) \label{lnvar} .
\end{align}
The average in the second term is readily computed,
\begin{align}
\Av{\big(\bm{x} - \bm{\mu}\big)^T \bm{\Xi}^{-1} \big(\bm{x} - \bm{\mu}\big)} = \Av{ \big(x_i-\mu_i \big) \big(\Xi^{-1}\big)_{i j} \big(x_j - \mu_j \big)} = \big(\Xi^{-1}\big)_{i j} \Av{ \big(x_i-\mu_i \big) \big(x_j - \mu_j \big)} = \big(\Xi^{-1}\big)_{i j} \Xi_{i j},
\end{align}
where summation over repeated indices is implied.
Since the covariance matrix is symmetric, this is equal to
\begin{align}
\Av{\big(\bm{x} - \bm{\mu}\big)^T \bm{\Xi}^{-1} \big(\bm{x} - \bm{\mu}\big)} = \big(\Xi^{-1}\big)_{i j} \Xi_{j i} = \text{Tr}(\bm{\Xi}^{-1} \bm{\Xi}) = \text{Tr}(\bm{1}) = M .
\end{align}
For the first term, on the other hand, we have
\begin{align}
\Av{\Big(\big(\bm{x} - \bm{\mu}\big)^T \bm{\Xi}^{-1} \big(\bm{x} - \bm{\mu}\big) \Big)^2} &= \Av{\big(x_i - \mu_i \big) \big(\Xi^{-1}\big)_{i j} \big(x_j - \mu_j \big) \big(x_k - \mu_k \big) \big(\Xi^{-1}\big)_{k l} \big(x_l - \mu_l \big)} \\
&= \big(\Xi^{-1}\big)_{i j} \big(\Xi^{-1}\big)_{k l} \Av{ \big(x_i - \mu_i \big) \big(x_j - \mu_j \big) \big(x_k - \mu_k \big) \big(x_l - \mu_l \big)} \n .
\end{align}
We now apply Isserli's theorem for higher order moments of normal random variables,
\begin{align}
\Av{ \big(x_i - \mu_i \big) \big(x_j - \mu_j \big) \big(x_k - \mu_k \big) \big(x_k - \mu_k \big)} = \Xi_{i j} \Xi_{k l} + \Xi_{i k} \Xi_{j l} + \Xi_{i l} \Xi_{j k}
\end{align}
and again use the symmetry of the covariance matrix to write,
\begin{align}
\big(\Xi^{-1}\big)_{i j} &\big(\Xi^{-1}\big)_{k l} \Av{ \big(x_i - \mu_i \big) \big(x_j - \mu_j \big) \big(x_k - \mu_k \big) \big(x_l - \mu_l \big)} \\
&= \big(\Xi^{-1}\big)_{i j} \Big( \Xi_{i j} \big(\Xi^{-1}\big)_{k l} \Xi_{l k} + \Xi_{i k} \big(\Xi^{-1}\big)_{k l} \Xi_{l j} + \Xi_{j k} \big(\Xi^{-1}\big)_{k l} \Xi_{l i} \Big) \n.
\end{align}
We now recast the sum over $l$ in matrix notation,
\begin{align}
\Xi_{i j} \big(\Xi^{-1}\big)_{k l} \Xi_{l k} + \Xi_{i k} \big(\Xi^{-1}\big)_{k l} \Xi_{l j} + \Xi_{j k} \big(\Xi^{-1}\big)_{k l} \Xi_{l i} &= \Xi_{i j} \big( \bm{\Xi}^{-1} \bm{\Xi} \big)_{k k} + \Xi_{i k} \big( \bm{\Xi}^{-1} \bm{\Xi} \big)_{k j} + \Xi_{j k} \big( \bm{\Xi}^{-1} \bm{\Xi} \big)_{k i} \\
&=\Xi_{i j} \delta_{k k} + \Xi_{i k} \delta_{k j} + \Xi_{j k} \delta_{k i} \nn
&=\Xi_{i j} M + \Xi_{i j} + \Xi_{j i} \n ,
\end{align}
where we performed the sum over $k$ in the last step.
We thus have
\begin{align}
\Av{\Big(\big(\bm{x} - \bm{\mu}\big)^T \bm{\Xi}^{-1} \big(\bm{x} - \bm{\mu}\big) \Big)^2} &= \big(\Xi^{-1}\big)_{i j} \Big( \Xi_{i j} M + \Xi_{i j} + \Xi_{j i} \Big) = M^2 + 2 M.
\end{align}
Plugging the results for the first and second term into Eq.~\eqref{lnvar}, we obtain the result,
\begin{align}
\Delta_{\ln} = \frac{M}{2} ,
\end{align}
independent of the form of the covariance matrix.

Next, for any distribution that depends on time only via its mean,
\begin{align}
P(\bm{x},t) = \tilde{P}(\bm{x} - \bm{\mu}(t)),
\end{align}
with a function $\tilde{P}(\bm{z})$ that does not explicitly depend on time, the Fisher information can be written as
\begin{align}
I(t) = \int d\bm{x} \ \frac{\big(\partial_t P(\bm{x},t)\big)^2}{P(\bm{x},t)} = \int d\bm{z} \ \frac{\big(\dot{\bm{\mu}}(t)^T \bm{\nabla}_z \tilde{P}(\bm{z})\big)^2}{\tilde{P}(\bm{z})} .
\end{align}
We now use the operator inequality, 
\begin{align}
\bm{D} - \bm{\Xi}^{-1} \geq 0 \label{operator-ineq},
\end{align}
in the sense that the expression on the left-hand side is a positive semidefinite matrix.
Here we defined
\begin{align}
(\bm{D})_{i j} = \int d\bm{z} \ \frac{\partial_{z_i}\tilde{P}(\bm{z}) \partial_{z_j}\tilde{P}(\bm{z}) }{\tilde{P}(\bm{z})} .
\end{align}
This inequality holds for arbitrary differentiable probability distributions and leads to
\begin{align}
I(t) = \dot{\bm{\mu}}(t)^T \bm{D} \dot{\bm{\mu}}(t) \geq \dot{\bm{\mu}}(t)^T \bm{\Xi}^{-1} \dot{\bm{\mu}}(t) .
\end{align}
Since the rightmost expression is just the Fisher information for a normal distribution with time-independent covariance matrix, Eq.~\eqref{fisher-normal}, this proves the bound \eqref{fisher-bound}.
What is left to do is to prove the operator inequality Eq.~\eqref{operator-ineq}.
To do so, we consider the covariance $\text{cov}(f,g) \equiv \av{f g}-\av{f} \av{g}$ with respect to some differentiable probability distribution $P(\bm{x})$, $\bm{x} \in \mathbb{R}^M$,
\begin{align}
\text{cov}(\bm{a}^T \bm{x}, \bm{b}^T \bm{\nabla} \ln(P)) &= \int d\bm{x} \ a_i x_i b_j \partial_{x_j} P(\bm{x}) - \int d\bm{x} \ a_i x_i P(\bm{x}) \int d\bm{y} \ b_j \partial_{y_j} P(\bm{x}) \\
& = -\int d\bm{x} \ a_i b_j  P(\bm{x}) \partial_{x_j} x_i = - a_i b_j \delta_{i j} \n ,
\end{align}
where $\bm{a}, \bm{b} \in \mathbb{R}^M$ are arbitrary vectors and we sum over repeated indices.
Here, we integrated by parts in the second-to-last step.
On the other hand, we have from the covariance inequality
\begin{align}
\text{cov}(\bm{a}^T \bm{x}, \bm{b}^T \bm{\nabla} \ln(P))^2 \leq \text{var}(\bm{a}^T \bm{x}) \text{var}(\bm{b}^T \bm{\nabla} \ln(P)) \label{cov},
\end{align}
where $\text{var}$ denotes the variance with respect to $P(\bm{x})$, $\text{var}(f) \equiv \av{f^2}-\av{f}^2$
First, we note that $\av{\bm{b}^T \bm{\nabla} \ln(P)} = 0$ and, consequently, the variance of $\bm{b}^T \bm{\nabla} \ln(P)$ is given by
\begin{align}
\text{var}(\bm{b}^T \bm{\nabla} \ln(P)) = \int d\bm{x} \ b_i \frac{\partial_{x_i} P(\bm{x}) \partial_{x_j} P(\bm{x})}{P(\bm{x})} b_j = \bm{b}^T \bm{D} \bm{b},
\end{align}
Next, we evaluate the variance
\begin{align}
\text{var}(\bm{a}^T \bm{x}) = \int d\bm{x} \ a_i a_j x_i x_j P(\bm{x}) - \int d\bm{x} \int d\bm{y} \ a_i a_j x_i y_j P(\bm{x}) P(\bm{y}) = \bm{a}^T \bm{\Xi} \bm{a}
\end{align}
Then, the covariance inequality \eqref{cov} can be written as
\begin{align}
\bm{b}^T \bm{D} \bm{b} \geq \frac{\big(\bm{b}^T \bm{a} \big)^2 }{\bm{a}^T \bm{\Xi} \bm{a}} .
\end{align}
Since this holds for arbitrary $\bm{a}$ and $\bm{\Xi}$ is positive definite and thus invertible, we may choose
\begin{align}
\bm{a} = \bm{\Xi}^{-1} \bm{b} .
\end{align}
For this choice, we obtain
\begin{align}
\bm{b}^T \bm{D} \bm{b} \geq \frac{\big(\bm{b}^T \bm{\Xi}^{-1} \bm{b} \big)^2 }{\bm{b}^T \bm{\Xi}^{-1} \bm{\Xi} \bm{\Xi}^{-1} \bm{b}} = \bm{b}^T \bm{\Xi}^{-1} \bm{b} ,
\end{align}
where we used the symmetry of $\bm{\Xi}$ and that $\bm{\Xi} \bm{\Xi}^{-1} = \bm{1}$.
Since $\bm{b}$ is arbitrary, this is equivalent to the inequality \eqref{operator-ineq}.

%A particular case is when we have $A(t) = \frac{\rho(t) - \rho(0)}{\sigma}$ with $\sigma$ independent of time and
%\begin{align}
%P_0(\bm{x}) = \frac{\exp\Big[{-\frac{(r(\bm{x})-\rho(0))^2}{2 \sigma}} \Big]}{\int d\bm{x} \ \exp\Big[{-\frac{(r(\bm{x})-\rho(0))^2}{2 \sigma}}\Big]}.
%\end{align}
%This corresponds to a pseudo-Gaussian distribution in terms of $r(\bm{x})$.
%Note that, unless $r(\bm{x})$ is a linear function $\rho$ and $\sigma$ are not the average and the variance of $r$, respectively, but rather parameters of the distribution.
%In this case, we find
%\begin{align}
%P(\bm{x},t) = \frac{\exp\Big[{-\frac{(r(\bm{x})-\rho(t))^2}{2 \sigma}} \Big]}{\int d\bm{x} \ \exp\Big[{-\frac{(r(\bm{x})-\rho(t))^2}{2 \sigma}}\Big]}.
%\end{align}
%The pseudo-Gaussian form of the distribution is thus preserved under the time evolution of the system, provided that the distribution depends on time only via the parameter $\rho(t)$.

\end{document}